\documentclass[11pt,a4paper]{article}
\usepackage{epsfig}
\usepackage[T1]{fontenc}    
\usepackage{graphics}
\usepackage{graphicx}
\usepackage{pstricks,pst-coil,pst-fill,pst-plot}
\usepackage[fleqn]{amsmath}    
\usepackage{amssymb}    
\usepackage{amsfonts}   
\usepackage{verbatim}   
\usepackage{mathrsfs}   
\usepackage{dsfont}
\usepackage{euscript}
\usepackage{yfonts}
\usepackage{txfonts}
\usepackage{marvosym}
\usepackage{vmargin}        
\usepackage{setspace}    

\setmarginsrb{1.8cm}{2cm}{1.8cm}{2cm}{1cm}{1cm}{1cm}{1.6cm}
 \makeatletter
 \@addtoreset{equation}{section}
 \makeatother

\providecommand{\bysame}{\leavevmode\hbox to3em{\hrulefill}\thinspace}

\let\tend=\rightarrow

\newtheorem{prop}{Proposition}[section]
\newtheorem{cor}{Corollary}[section]
\newtheorem{defin}{Definition}[section]

\newtheorem{lemme}{Lemma}[section]

\def\Proof{\medskip\noindent {\it Proof --- \ }}

\def\qed{\hfill\rule{2mm}{2mm}}

\newcommand\beq{\begin{equation}}
\newcommand\enq{\end{equation}}
\newcommand\bem{\begin{multline}}
\newcommand\enm{\end{multline}}

\def\beqa{\begin{eqnarray}}
\def\eeqa{\end{eqnarray}}
\def\ba{\begin{array}}
\def\ea{\end{array}}

\newcommand{\f}[2]{{\ensuremath{%
    \mathchoice%
    {\dfrac{#1}{#2}}
    {\dfrac{#1}{#2}}
    {\frac{#1}{#2}}
    {\frac{#1}{#2}}
}}}

\newcommand{\pa}[1]{\ensuremath{\left(#1\right)}}

\newcommand{\pac}[1]{\ensuremath{\left[#1\right]}}

\def\a{\alpha}

\def\ga{\gamma}
\def\Ga{\Gamma}
\def\de{\delta}

\def\eps{\epsilon}
\def\veps{\varepsilon}
\def\la{\lambda}

\def\sg{\sigma}

\def\Om{\Omega}
\def\om{\omega}
\def\vp{\varphi}

\newcommand{\mc}[1]{\ensuremath{\mathcal{#1}}}
\newcommand{\mf}[1]{\ensuremath{\mathfrak{#1}}}
\newcommand{\msc}[1]{\ensuremath{\mathscr{#1}}}

\newcommand{\bs}[1]{\ensuremath{\boldsymbol{#1}}}

\newcommand{\ov}[1]{\ensuremath{\overline{#1}}}
\newcommand{\wt}[1]{\ensuremath{\widetilde{#1}}}
\newcommand{\wh}[1]{\ensuremath{\widehat{#1}}}

\newcommand{\Int}[2]{\ensuremath{\int\limits_{#1}^{#2}}}
\newcommand{\Oint}[2]{\ensuremath{\oint\limits_{#1}^{#2}}}

\newcommand{\sul}[2]{\ensuremath{\sum\limits_{#1}^{#2}}}
\newcommand{\pl}[2]{\ensuremath{\prod\limits_{#1}^{#2}}}

\newcommand{\R}{\ensuremath{\mathbb{R}}}
\newcommand{\Cx}{\ensuremath{\mathbb{C}}}

\newcommand{\Dp}[1]{\ensuremath{\partial_{#1}}}

\newcommand{\limit}[2]{\ensuremath{\underset{#1 \tend #2}{\longrightarrow} }}

\newcommand{\ex}[1]{\ensuremath{\e{e}^{#1}}}

\newcommand{\bra}[1]{\left\langle \,#1\,\right|}
\newcommand{\ket}[1]{\left|\,#1\, \right\rangle}

\newcommand{\norm}[1]{\ensuremath{ || #1 || }}

\newcommand{\dd}{\mathrm{d}}
\newcommand{\e}[1]{\ensuremath{\mathrm{#1}}}

\newcommand{\intn}[2]{\ensuremath{[\![ \, #1 \,;\, #2 \,]\!]}}

\begin{document}

\begin{flushright}

\end{flushright}
\par \vskip .1in \noindent

\vspace{14pt}

\begin{center}
\begin{LARGE}
{\bf Aspects of the inverse problem for the Toda chain.}
\end{LARGE}

\vspace{30pt}

\begin{large}

{\bf K.~K.~Kozlowski}\footnote[1]{Universit\'{e} de Bourgogne, Institut de Math\'{e}matiques de Bourgogne, UMR 5584 du CNRS, France,
karol.kozlowski@u-bourgogne.fr}. 
%

\end{large}

\vspace{40pt}

\centerline{\bf Abstract} \vspace{1cm}
\parbox{12cm}{\small 
We generalize Babelon's approach to equations in dual variables so as to be able to treat
new types of operators which we build out of the sub-constituents of the model's monodromy matrix. 
Further, we also apply Sklyanin's recent monodromy
matrix identities so as to obtain equations in dual variables for yet other operators. 
The schemes discussed in this paper appear to be universal and thus, in principle, applicable to many models
solvable through the quantum separation of variables. 
}

\end{center}

\vspace{40pt}

\section*{Introduction}

\label{INT}

The Toda chain refers to a quantum mechanical $N+1$-body Hamiltonian in one spatial dimension 
\beq
\bs{ H }_{\kappa} \; = \;  \sul{a=1}{N+1} \f{p_{a}^2}{2} \; + \;\kappa \ex{ x_{N+1} - x_1} 
\; + \;  \sul{a=1}{N} \ex{x_a -x_{a+1} } \qquad \e{with} \quad p_n = \f{\hbar}{i} \f{ \Dp{} }{ \Dp{} x_n } \;. 
\enq
There, $p_n$ and $x_n$ are pairs of conjugated variables satisfying the canonical commutation $\pac{x_k,p_{\ell}} = i\hbar$. 
Also, the index $n$ refers to the quantum space where these operators act non-trivially. 
When $ \kappa =1$, one deals with the so-called closed Toda chain whereas the model at $\kappa=0$ is referred to
as the open Toda chain. 

The classical counterpart of the model has been introduced by Toda \cite{TodaIntroTodaAndClassicalSolutionTodaChain}.
Its classical integrability has been established in \cite{FlaschkaLaxMatrixIntegrabilityClassicalToda,KacMoerbeckeFullSolutionTodaAbelianIntegrals}. 
Explicit formulae for the inverse action-angle map have been obtained first by Ruijssenaars \cite{RuijsenaarsActionAngleMapsForVeriousSystems} and 
recently rederived by Feh\'{e}r \cite{FeherActionAngleMapOpenClassicalToda} by means of a much simpler setting. 
Further, 
Olshanetsky and Perelomov \cite{OlshanetskyPerelomovIntegralsOfMotionSemi-SimpleLieAlgebraSystems} constructed the quantum integrals of motion inductively whereas Kostant \cite{KostantIdentificationOfEigenfunctionsOpenTodaAndWhittakerVectors}
identified eigenfunctions of the open chain with Whittaker functions for $GL(N,\R)$. 
The explicit characterization of the spectrum  of $\bs{H}_{\kappa=1}$ has been first investigated 
by Gutzwiller \cite{GutzwillerResolutionTodaChainSmallNPaper1,GutzwillerResolutionTodaChainSmallNPaper2} for small
values of $N$ (N=1, 2, 3) through a direct analysis of the partial
differential equation. 
His main achievement was to express the eigenfunctions of the $N+1$-body periodic
chain in terms of an integral transform whose kernel corresponds to the generalized eigenfunctions of the $N$-body
open Toda chain. This integral transform also involved a function solving a second order difference equation in 
one variable, the $T-Q$ equation \cite{BaxterPartitionfunction8Vertex-FreeEnergy}.  Gaudin and Pasquier  
were the first to obtain the operator valued $T-Q$ equations associated with this model, this for any value of $N$. 
Then, Sklyanin \cite{SklyaninSoVFirstIntroTodaChain} introduced the so-called quantum separation of variables
what allowed him to derive the scalar form of the $T-Q$ equations for the Toda chain this, as well, for any
value of $N$. In fact, the quantum separation of variables is realised by means of an integral transform. 
Namely, define the transform 
\beq
\Phi \big( \bs{x}_{N+1} \big)  \; = \; \Int{\R^{N+1} }{} \Psi_{\bs{y}_{N};\veps} (\bs{x}_{N+1})  
\wh{\Phi}(\bs{y}_{N}; \veps ) \,  \cdot \,   \f{ \dd\mu(\bs{y}_{N}) }{ \sqrt{N!} }  \otimes \dd \veps \;, 
\label{ecriture forme transfor int Sov pour Intro}
\enq
where the subscript indicates the dimensionality of the vectors, \textit{ie} 
$\bs{x}_{N+1}=(x_1,\dots, x_{N+1})$ and $\bs{y}_N = ( y_1 , \dots, y_N)$. Finally, $\dd\mu(\bs{y}_{N})$
is the Sklyanin measure. The purpose of this transform is to map the multidimensional spectral problem
associated with $\bs{H}_{\kappa=1}$ onto a one dimensional spectral problem. It is in this sense that 
one speaks of separation of variables.   

As observed by Gutzwiller, the correct object for defining the kernel of the integral transform are
eigenfunctions $\vp_{\bs{y}_N}(\bs{x}_N)$ of the open Toda chain -$\bs{H}_{\kappa=0}$- with $N$-particles. Namely one should take
\beq
\Psi_{\bs{y}_{N};\veps} (\bs{x}_{N+1}) \; = \;  \ex{\f{i}{\hbar} ( \eps \,  - \, \ov{\bs{y}}_N ) x_{N+1} } \cdot
 \vp_{ \bs{y}_N }\big(  x_N \big)\;. 
\enq

The map $\wh{ \Phi} \mapsto \Phi$, defined
on $L^{1}_{ \e{sym} \times - }\big( \R^N \times \R ,  \dd\mu(\bs{y}_N) \otimes \dd \veps \big)$,
 extends to an unitary map from $L^{2}_{ \e{sym} \times - }\big( \R^N \times \R ,  \dd\mu(\bs{y}_N) \otimes \dd \veps \big)$ onto 
$L^{2}\big( \R^{N+1} , \dd^{N+1} x)$. 
There, the subscript $_{ \e{sym} \times - }$
indicates that the functions are symmetric in the first set of $N$ variables.
This unitarity has been first established, within group theoretical based arguments. 
The work \cite{SemenovTianShanskyQuantOpenTodaLatticesProofOrthogonalityFormulaForWhittVectrs} proved the orthogonality 
that is to say the isometric character of the inverse transform whereas completeness follows
from the arguments that can be found in \cite{WallachRealReductiveGroupsII}. Then, in \cite{SilantyevScalarProductFormulaTodaChain},
a formal proof, based on techniques developed in \cite{DerkachovKorchemskyManashovXXXSoVandQopNewConstEigenfctsBOp},
of the orthogonality of the transform has been proposed. Finally, the author \cite{KozUnitarityofSoVTransform}
gave recently a new proof of the transform's unitarity. The proof given by the author was based, on the one hand, on bringing
rigour to the arguments of \cite{SilantyevScalarProductFormulaTodaChain} and, on the other hand, in developing a 
new technique allowing one to prove completeness, this solely by using the quantum inverse scattering framework. 
Unitarity being established, the characterization of the spectrum boils down to solving the model's $T-Q$ scalar equations
as shown by An \cite{AnCompletenessEigenfunctionsTodaPeriodic}. 
The latter's solution can be described either on the algebraic 
\cite{GaudinPasquierQOpConstructionForTodaChain,GutzwillerResolutionTodaChainSmallNPaper1,GutzwillerResolutionTodaChainSmallNPaper2} 
or non-linear integral equation \cite{KozTeschnerTBAToda,NekrasovShatashviliConjectureTBADescriptionSpectrumIntModels} 
levels quite explicitly. 
Hence, it is quite fair to state that,  as of today, the understanding of the structure of the space of states and of the 
model's spectrum are quite good.  

However, from the perspective of applications to physics, it is the access to a model's correlation functions that is the most 
interesting. Taking into account the natural simple description of the closed Toda chain's
eigenfunction on the space $L^{2}_{ \e{sym}\times - }\big( \R^N \times \R ,  \dd\mu(\bs{y}_N) \otimes \dd \veps \big)$,
it appears most reasonable to compute correlation functions by solving the so-called inverse problem, \textit{ie}. 
compute the expectation values $\bra{\Phi_1} \mc{O} \ket{\Phi_2}$ by passing to the representation 
of the model's Hilbert space on 
$L^{2}_{ \e{sym}\times - }\big( \R^N \times \R ,  \dd\mu(\bs{y}_N) \otimes \dd \veps \big)$. This operation 
means that one should manage to express how operators $\mc{O}$ having a simple (\textit{ie}. local) 
action on the model's original space act directly on the space of functions where the separation of variables occurs. 
Due to the structure of the transform \eqref{ecriture forme transfor int Sov pour Intro} it is, in fact, enough 
to determine how the action of such operators translates itself on the dual ($\bs{y}_N; \veps$) 
variables of the kernel functions $\Psi_{\bs{y}_{N};\veps} (\bs{x}_{N+1})$, \textit{ie}. obtain an equation
\beq
\mc{O} \cdot \Psi_{\bs{y}_{N};\veps} (\bs{x}_{N+1}) \; = \; \wh{\mc{O}} \cdot  \Psi_{\bs{y}_{N};\veps} (\bs{x}_{N+1})
\enq
in which the operator $\mc{O} $ acts on the space variables  $\bs{x}_{N+1}$ whereas its dual operator $\wh{\mc{O}}$
acts on the dual ones $\bs{y}_N; \veps$.

The inverse problem for integrable models solvable by the algebraic Bethe Ansatz method has been first
solved in \cite{KMTFormfactorsperiodicXXZ} and further developed in  
\cite{MailletTerrasGeneralsolutionInverseProblem,OotaInverseProblemForFieldTheoriesIntegrability}. 
One can in fact say that, within today's state of the art, the resolution of the inverse problem for models
solvable by the algebraic Bethe Ansatz is quite well understood. The situation is however not so well established in what concerns models 
solvable by the quantum separation of variables method. 
In \cite{BabelonQuantumInverseProblemConjClosedToda}, Babelon derived, on the basis of semi-classical 
arguments, the form certain local operators associated with the Toda chain take on 
$L^{2}_{ \e{sym}\times - }\big( \R^N \times \R ,  \dd\mu(\bs{y}_N) \otimes \dd \veps \big)$. He then justified \cite{BabelonActionPositionOpsWhittakerFctions} 
one set of his formulae by computing the action of these operators on Whittaker functions. 
Recently, Sklyanin \cite{SklyaninResolutionIPFromQDet} managed to reproduce Babelon's formulae through simple algebraic 
arguments based on the quantum inverse scattering approach to the quantum Toda chain. 

A different route to solving inverse problems for certain quantum separation of variables
models - those associated with finite dimensional representations attached to each lattice node - 
has been proposed in  \cite{GrosjeanMailletNiccoliFFofLatticeSineG}. This method builds on Oota's 
\cite{OotaInverseProblemForFieldTheoriesIntegrability} ideas for solving the inverse problem for algebraic Bethe Ansatz solvable models 
as well as on certain properties associated with the finite dimensionality of the representations. 
It was applied to other models in a subsequent series of works, see \textit{eg}. \cite{NiccoliCompleteSpectrumAndSomeFormFactorsInhomogeneousOpenXXZChain}. 
However, the method works only, \textit{per se}, for inhomogeneous deformations of an integrable model
of interest. Although, within such an approach, the final expression for the correlation
functions have a well-defined homogeneous limit, its characterisation in a convenient form 
still remain an open problem.  

In the present paper, we push forward the techniques developed by Babelon
\cite{BabelonActionPositionOpsWhittakerFctions,BabelonQuantumInverseProblemConjClosedToda}
and demonstrate that one can derive equations in dual variables for more general operators. 
Such operators are built out of certain sub-components of the model's monodromy matrix. 
Due to the natural quantum inverse scattering method interpretation of these operators, 
we believe that our construction can be generalized to other, more complex models. 
Furthermore, in fact, we show that Sklyanin's recent observations \cite{SklyaninResolutionIPFromQDet} 
allow one to obtain equations in dual variables for an even larger class of operators.

The paper is organised as follows. In Section \ref{Section integrabilite chaine Toda rappels}, we revisit
certain aspects of the quantum integrability of the Toda chain. After recalling the main 
ingredients of the quantum inverse scattering method approach to this model, we build new types of 
Mellin-Barnes multiple integral representations for the function $\Psi_{\bs{y}_{N};\veps} (\bs{x}_{N+1})$. 
In Section \ref{Section resolution probleme inverse}, we gather the main results of this paper, namely,
a set of equations in dual variables for certain classes of operators built out of sub-components of the 
monodromy matrix. Most of the proofs and technical details are gathered in three appendices.

\section{Integrability of the quantum Toda chain}
\label{Section integrabilite chaine Toda rappels}

\subsection{The Lax matrix formulation}

The quantum integrability of the Toda chain can be described within the framework of the quantum inverse scattering 
method. The central object in this approach is a $2\times 2$ Lax matrix 
\beq
L_{0n}(\la) \; = \; \pa{\ba{cc} \la-p_n & \ex{-x_n} \\ - \ex{x_n} & 0 \ea}_{\pac{0}} \quad \e{with} \quad 
 \pac{x_k,p_{\ell}} = i\hbar \de_{k\ell} \;.
\label{ecriture matrice de Lax Toda}
\enq

It is straightforward to check that the latter satisfies the below quadratic algebra
\beq
R_{00^{\prime}}(\la-\mu) L_{0n}(\la) L_{0^{\prime} n}(\mu) \; = \; L_{0^{\prime} n}(\mu) L_{0n}(\la) R_{00^{\prime}}(\la-\mu) 
\label{ecriture YBE local matrice de Lax}
\enq
where the $4\times 4$ $R$-matrix reads:
\beq
R_{00^{\prime}}(\la) \; = \; \left( \ba{cccc} \la+i\hbar &  0  & 0 & 0 \\
												0   &   \la  & i\hbar  & 0 \\
												0 & i \hbar & \la & 0  \\ 
												0  & 0 & 0 & \la+i\hbar \ea \right).  
\enq
Out of such matrices, one builds the so-called monodromy matrix of the model as an ordered product of local Lax matrices:
\beq
T_{0;1,N+1}(\la)  \; = \ L_{01}(\la) \dots L_{0N+1}(\la) \;  =  \; 
	\pa{ \ba{cc} A_{1, N+1} \pa{\la} & B_{1, N+1} \pa{\la} \\ 
	C_{1, N+1} \pa{\la} & D_{1, N+1}\pa{\la} \ea}_{\pac{0}}\;.
\enq

The ultra-local algebra \eqref{ecriture YBE local matrice de Lax} satisfied 
by the Lax matrices raises to a quadratic algebra, the so-called Yang--Baxter algebra, relating the entries of the monodromy matrix:
\beq
R_{00^{\prime}}(\la-\mu) T_{0;1,N+1}(\la) T_{0^{\prime};1,N+1}(\mu)  \; = \; 
T_{0^{\prime};1,N+1}(\mu) T_{0;1,N+1}(\la)   R_{00^{\prime}}(\la-\mu)  \;. 
\label{ecriture YBE pour matrice de monodromie}
\enq
The relations encoded in the above algebra are sufficiently rich so as to provide 
one with the full spectrum and complete set of eigenfunctions of the $N$-body closed Toda chain. The simplest, yet by no means 
less important, consequence of the above Yang-Baxter algebra for the monodromy matrix is the possibility to provide a set of $N+1$
Hamiltonians in involution, which, in particular, contains $\bs{H}_{\mid \kappa=1}$. 
In order to do so, one defines the so-called transfer matrix $ \bs{\tau}(\la) \; = \; \e{tr}_0\big[ T_{0;1,N+1}(\la)   \big]$
of the model. The Yang-Baxter equation ensures that $\bs{\tau}(\la)$ 
gives rise to a one parameter $\la$ commutative subalgebra of operators on $L^2(\R^{N+1}, \dd^{N+1}x)$. 
Since $\bs{\tau}(\la)$ is a monoic operator valued polynomial in $\la$ of degree $N+1$, the transfer matrix gives rise to
a set of $N+1$ Hamiltonians in involution. These can be, for instance, defined as the coefficients arizing in the 
$\la$-expansion of $ \bs{\tau}(\la) $:
\beq
\bs{\tau}(\la) \; = \; \la^{N+1}  \; + \; \sul{k=1}{N+1}(-1)^k \la^{N-k+1} \bs{\tau}_k \qquad \e{with} \qquad 
\bs{\tau}_1 \; = \; \sul{a=1}{N+1} p_a \qquad \e{and}  \qquad  \bs{\tau}_2 \; = \; \bs{\tau}_1^2 - \bs{H}_{\mid \kappa=1} \;. 
\enq

The Lax matrix given in \eqref{ecriture matrice de Lax Toda}  
				can be explicitly inverted with the help of the quantum determinants relation
\beq
\big[  L_{0n}(\la) \big]^{-1} \; = \; \left( \ba{cc} 0 & -\ex{-x_n} \\ 
										\ex{x_n}  & \la - i \hbar -p_n  \ea \right)_{[0]} 
\; = \;  \sg_0^y \cdot L_{0n}^{t_0}(\la-i\hbar)	\cdot \sg_0^y  	\;. 
\label{formule inverse op de Lax}
\enq
Just as in the case of the local Yang--Baxter algebra, this local inversion formula can be raised to the level 
of the monodromy matrix leading to 
\beq
\big[ T_{0;1,N+1}(\la) \big]^{-1}  \; = \; \sg_0^y T_{0;1,N+1}^{t_0}(\la-i\hbar) \sg_0^y  \; . 
\enq
This explicit realisation for the inverse of the monodromy matrix leads to the so-called quantum determinant relations. 
By reading out the various entries of the matrix product in the identity 
$   \big[ T_{0;1,N+1}(\la) \big] \cdot \big[ T_{0;1,N+1}(\la) \big]^{-1}   \; = \; I_2 \otimes \e{id}_{L^2(\R^N)}$ 
one obtains 
\beq
\left\{  \ba{ccc} A_{1,  N +1 }(\la) \, D_{1,  N + 1}(\la-i\hbar) \; - \; B_{1,  N+1 }(\la) \, C_{1,  N +1}(\la-i\hbar) 
& = &  \e{id}_{L^2(\R^{N+1})} \\
D_{1,  N +1 }(\la) \, A_{1,  N + 1}(\la-i\hbar) \; - \; C_{1,  N+1 }(\la) \, B_{1,  N +1}(\la-i\hbar) 
& = &  \e{id}_{L^2(\R^{N+1})}   \ea \right. 
\label{ecriture qDet relation egalite a identite}
\enq
and also 
\beq
\left\{  \ba{ccc}  
			C_{1,  N +1 }(\la) \, D_{1,  N + 1}(\la-i\hbar) & = & D_{1,  N+1 }(\la) \, C_{1,  N +1}(\la-i\hbar) \\
B_{1,  N +1 }(\la) \, A_{1,  N + 1}(\la-i\hbar) & = & A_{1,  N+1 }(\la) \, B_{1,  N +1}(\la-i\hbar) \ea \right. 
\; . 
\label{ecriture qDet relation relation commutation}
\enq

It is readily seen that $B_{1,N+1}\pa{\la} = \ex{-x_{N+1}} A_{1,N}\!\pa{\la}$ so that 
$B_{1,N+1}\pa{\la}$ is realized as a multiplication operator 
in respect to $x_{N+1}$. Furthermore, the operator $A_{1,N}\!\pa{\la}$ is the generating function of the
integrals of motion for the open $N$-particle Toda chain. 

In fact, one way of defining the integral kernel $\Psi_{\bs{y}_{N};\veps} (\bs{x}_{N+1})$ of the SoV transform 
\eqref{ecriture forme transfor int Sov pour Intro} is as the functions satisfying to the equation
\beq
B_{1,N+1}(\la) \Psi_{\bs{y}_{N},\veps} (\bs{x}_{N+1})\; = \; 
\pl{a=1}{N}(\la-y_a) \cdot \Psi_{\bs{y}_{N},\veps+i\hbar} (\bs{x}_{N+1}) \;. 
\label{ecriture equation definition heuristique des fct propres B}
\enq
The latter equation, along with a multiplicity one theorem \cite{GoodmannWallachQuantumTodaIII,KozUnitarityofSoVTransform,
SemenovTianShanskyQuantOpenTodaLatticesProofOrthogonalityFormulaForWhittVectrs,WallachRealReductiveGroupsII}, \textit{ie} the completeness and 
orthogonality of the system $\Psi_{\bs{y}_{N};\veps} (\bs{x}_{N+1})$, completely 
characterises these functions. 

In order to establish the equations in dual variables for $\Psi_{\bs{y}_{N};\veps} (\bs{x}_{N+1})$
that are at the core of the present paper, we shall construct a generalization of the Mellin-Barnes multiple integral representation 
for $\Psi_{\bs{y}_{N};\veps} (\bs{x}_{N+1})$ that has been initially obtained by Kharchev and Lebedev
\cite{KharchevLebedevMellinBarnesIntRepForWhittakerGLN,KharchevLebedevIntRepEigenfctsPeriodicTodaFromRecConstrofEigenFctOfB}
following the procedure suggested by Sklyanin in \cite{SklyaninSoVGeneralOverviewAndConstrRecVectPofB}. 
More precisely, the Mellin-Barnes integral representation is obtained recursively by demanding that equation 
\eqref{ecriture equation definition heuristique des fct propres B} holds and acting with the $B_{1,N+1}$-operator 
by means of its decomposition associated with the splitting of the original chain of length
$N+1$ into two-sub chains of respective lengths $r$ and $N+1-r$. The case originally dealt with by Kharchev and Lebedev 
corresponds to $r=1$. However many properties of the Whittaker functions and in particular the equations in the dual variables
are more apparent to see from representations subordinate to a splitting with a general $r$. 
Indeed, as observed by Babelon \cite{BabelonActionPositionOpsWhittakerFctions}, the Mellin-Barnes representation 
is perfectly adapted for solving the inverse problem. The main technical reason for this is that this
multiple integral representation, as opposed to the Gauss--Givental one \cite{GiventalGaussGivIntRepObtainedForEFOfOpenToda}, 
does not treat all the entries of the vector $\bs{x}_{N+1}$ on an equal footing 
but rather singles out a portion thereof.

\subsection{Inductive construction of the Mellin-Barnes representation and its basic properties}

It will be convenient, in the following, to introduce the notation $\ov{\bs{z}}_k$, which means that for any 
$k$-dimensional vector $\bs{z}_k=(z_1,\dots,z_k)$, $k \in \mathbb{N}$,
\beq
\ov{\bs{z}}_k \; = \; \sul{s=1}{k} z_s \; . 
\enq
Sometimes, when there will be no ambiguity on the dimensionality of the vector, we will simply drop 
the associated subscript, \textit{eg}.
\beq
\ov{\bs{y}} \; =  \; \sul{p=1}{N} y_p \;. 
\enq

\begin{defin}
\label{Definition procedure recursive construction fcts propre B}
Let $\bs{y}_N \in \Cx^N$ and $\veps \in \Cx$ be given. Then, define 
the function $\Psi_{\bs{y}_N,\veps}(\bs{x}_{N+1})$ inductively as follows. 
First set 
\beq
\Psi_{\emptyset , \veps} (x)  \; = \; \ex{i\f{\veps}{\hbar} x}
\enq
and then define the collection of functions $\Psi_{\bs{y}_N, \veps} (\bs{x}_{N+1})$ by 
\beq
\Psi_{\bs{y}_N, \veps} (\bs{x}_{N+1})  \; = \; \Int{ \msc{C}_{r-1; N-r} }{}  \Psi_{ \bs{w} ; \ov{\bs{y}}_N-\ov{\bs{z}} }(\bs{x}_1) 
\Psi_{\bs{z} ; \veps-\ov{\bs{y}}_N+\ov{\bs{z}} }(\bs{x}_2) \;  
\varpi(\bs{w},  \bs{z} \mid \bs{y}_N) \cdot \pl{a=1}{r-1}\dd w_a  \cdot \pl{a=r}{N-1}\dd z_a  \;. 
\label{ecriture rep int fct propre B}
\enq
There the integration runs through two sets of variables $w_a$ and $z_b$ which are collected in a vector notation
\beq
\bs{w} \; = \; (\underbrace{w_1,\dots, w_{r-1}, 0, \dots, 0}_{N-1})  \qquad \e{and} \qquad 
\bs{z} \; = \; (\underbrace{0,\dots, 0, z_r,\dots, z_{N-1}}_{N-1}) \; .
\label{definition des deux ensembles de variables d'integration}
\enq
The vectors 
\beq
\bs{x}_1 = (x_1,\dots, x_{r} )  \qquad \e{and} \qquad 
			\bs{x}_2 = (x_{r+1},\dots, x_{N+1} )
\enq
correspond to a splitting of the coordinates of the position vector $\bs{x}_{N+1} = (x_1,\dots, x_{N+1} )$. 
Further, the integration runs through the domain
\beq
\msc{C}_{r-1; N-r} \; = \; \big(\R-i\a\big)^{r-1}\!\! \times \big(\R+i\a\big)^{N-r} \quad \e{where} \; \a 
\; \e{is} \; \e{such} \; \e{that} \quad 
 \a > \max_{k\in \intn{1}{N}} \big( |\Im(y_k)| \big) \;. 
\label{definition du contour d'intégration pour fcton Whittaker}
\enq
Finally, the weight $\varpi(\bs{w},  \bs{z} \mid \bs{y}_N)$ arising under the integral sign is given by  
\beq
\varpi(\bs{w},  \bs{z} \mid \bs{y}_N) \; = \; \f{ (2\pi \hbar)^{1-N} }{(r-1)! (N-r)! } \cdot 
\f{ \pl{a=1}{r-1} \pl{b=1}{N} \bigg\{  \Ga\Big( \f{y_b-w_a}{ i \hbar} \Big) \, \hbar^{\f{i}{\hbar} (w_a-y_b)}  \bigg\}  
\cdot \pl{a=r}{N-1} \pl{b=1}{N} \bigg\{  \Ga\Big( \f{z_a-y_b}{ i \hbar} \Big) \,  \hbar^{\f{i}{\hbar} (y_b-z_a)}  \bigg\}  }
{  \pl{ \substack{a, b =1 \\a \not= b} }{ r-1} \Ga\Big( \f{w_a-w_b}{ i \hbar} \Big) \cdot 
\pl{ \substack{a,b=r \\ a \not= b} }{ N-1} \Ga\Big( \f{z_a-z_b}{ i \hbar} \Big) 
\cdot \pl{a=1}{r-1} \pl{b=r}{N-1} \bigg\{  \Ga\Big( \f{z_b-w_a}{ i \hbar} \Big) \,  \hbar^{\f{i}{\hbar} (w_a-z_b)}  \bigg\}  } \;. 
\enq
\end{defin}

\noindent There are several points that ought to be addressed in respect to this definition. 
\begin{itemize}
\item One should check that it is clean-cut, \textit{ie}. that the integrals \eqref{ecriture rep int fct propre B} are 
indeed convergent;
\item one should establish the natural properties of the functions defined through \eqref{ecriture rep int fct propre B} such as their asymptotic behaviour in the $y_a$'s and their regularity 
properties in respect to the variables $(\bs{y}_N, \veps )$ and $\bs{x}_{N+1}$; 
\item one should establish that it is consistent, \textit{ie}. that the functions 
$\Psi_{\bs{y}_N;\veps}(\bs{x}_{N+1})$ do not depend on the value of the integer $r$ used in the splitting of the 
integration variables \eqref{definition des deux ensembles de variables d'integration}.
\end{itemize}
We shall investigate the first two points in the proposition below.

\begin{prop}
\label{Proposition Estimates for decay at infinity of eigenfunction B}

Let $\eta>0$ be fixed and such that $|\Im(y_k)| <\eta$ for all $k=1,\dots,N$ as well as $|\Im(\veps)|<\eta$. 
Then, given  $N$, there exist constants $C_1$, $C_2$ and $M$ such that the 
functions defined through \eqref{ecriture rep int fct propre B} satisfy to the bounds 
\beq
\big|  \Psi_{\bs{y}_N;\veps}(\bs{x}_{N+1}) \big|  \; \leq  \;  C_1 \ex{ C_2 \norm{\bs{x}_{N+1}} }  \cdot 
\big( 1+ \norm{\bs{y}_N} \big)^{M}  \cdot  \exp\Big\{ -\f{\pi}{2\hbar} \sul{a<b}{N} |\Re(y_a-y_b)|  \;  \Big\} \;. 
\label{ecriture estimation de fction Psi par rapport  vars y}
\enq
In particular, the multiple integral -independently of the value of $r$- in \eqref{ecriture rep int fct propre B} is convergent and defines an entire functions 
of $(\bs{y}_N,\eps, \bs{x}_{N+1})$.  
Note that $\norm{ \cdot }$  refers to the $L^{1}$ norm on $\R^k$, with $k$-being the dimensionality of the vector, namely,
for a $p$-dimensional vector $\bs{v}_p$, one has 
\beq
\norm{\bs{v}_p} \; = \; \sul{a=1}{p} |v_a|  \;. 
\label{definition normes L1 vecteurs}
\enq
Finally, the constants $C_1, C_2$ in \eqref{ecriture estimation de fction Psi par rapport  vars y}
do not depend on the chain of splitings $r$ used in the application of the Definition \ref{Definition procedure recursive construction fcts propre B} so as to recursively build the function $\Psi_{\bs{y}_N;\veps}(\bs{x}_{N+1}) $.  

\end{prop}

\noindent We postpone the proof of Proposition \ref{Proposition Estimates for decay at infinity of eigenfunction B} to appendix \ref{Appendix Proof prop decay eigenfcts B op} and continue by listing the properties of the $\Psi_{\bs{y}_N,\veps}(\bs{x}_{N+1})$
functions in respect to the action of matrix entries of the monodromy matrix as well as by establishing the 
independence on the splitting $r$ of the whole construction.

\begin{prop}
\label{propositon action operateurs ABD}

The part of construction \eqref{ecriture rep int fct propre B} for the function $\Psi_{\bs{y}_N,\veps}(\bs{x}_{N+1})$
does not depend on $r$ and the latter function satisfies to the identities: 
\beq
\big[B_{1,   N+1}(\la) \Psi_{\bs{y}_N, \veps} \big] (\bs{x}_{N+1}) \; = \; 
			\pl{a=1}{N} \big(\la-y_a) \cdot \Psi_{ \bs{y}_N, \veps+i\hbar } (\bs{x}_{N+1}) \;. 
\label{ecriture action B sur Psi}
\enq
Also, $\Psi_{\bs{y}_N,\veps}$ fulfils 
\beq
\big[ D_{1,N+1}(\la) \Psi_{\bs{y}_N, \veps} \big] (\bs{x}_{N+1}) \; = \; 
\sul{p=1}{N} (-i)^{N+1} \pl{ \substack{ s=1 \\ \not=p} }{ N } \Big( \f{\la-y_s}{y_p-y_s} \Big) 
\cdot  \Psi_{\bs{y}_N-i\hbar \bs{e}_p, \veps} (\bs{x}_{N+1}) \;, 
\label{ecriture action D sur Psi}
\enq
as well as
\beq
\big[ A_{1,N+1}(\la) \Psi_{\bs{y}_N, \veps} \big] (\bs{x}_{N+1}) 
\; = \; \big( \la-\veps + \ov{\bs{y}}_N \big)  \pl{\ell=1}{N}(\la-y_{\ell}) \Psi_{\bs{y}_N, \veps} (\bs{x}_{N+1})
\; + \; \sul{p=1}{N} (i)^{N+1} \pl{ \substack{ s=1 \\ \not=p} }{ N } \Big( \f{\la-y_s}{y_p-y_s} \Big) 
\cdot  \Psi_{\bs{y}_N+i\hbar \bs{e}_p, \veps} (\bs{x}_{N+1}) \;. 
\label{ecriture action A sur Psi}
\enq
There $\bs{e}_p$ stands for the unit vector in $\R^N$ with a $1$ solely in its $p^{\e{th}}$ entry:
\beq
\bs{e}_p \; = \; \big( \underbrace{0, \dots, 0}_{p-1 \;  \e{terms} } , 1, 0,\dots, 0  \big) \;. 
\enq
Finally, one also has the identities
\beq
\pl{a=1}{N} \ex{x_a} \cdot \Psi_{\bs{y}_N,\veps}(\bs{x}_{N+1}) \; = \; 
 \Psi_{\bs{y}_N-i\hbar \bs{e},\veps- i \hbar N}(\bs{x}_{N+1}) \qquad \e{and}  \qquad
\ex{-\ell x_{N+1}} \cdot \Psi_{\bs{y}_N,\veps}(\bs{x}_{N+1}) \; = \; \Psi_{\bs{y}_N,\veps+i\ell \hbar}(\bs{x}_{N+1}) \;. 
\label{ecriture proprietes action ex x sur fct propre B}
\enq
The vector $\bs{e}$ introduced above takes the form $\bs{e} = \sul{p=1}{N} \bs{e}_p$ .

\end{prop}

The proof of the proposition is postponed to Appendix \ref{Appendix Proof Prop action ABD}

We remind that the action of the $C_{1,\dots,N+1}(\la)$ operator on $\Psi_{\bs{y}_N, \veps} (\bs{x}_{N+1})$ 
can be obtained from the quantum determinant relation
\beq
C_{1,  N+1}(\la) \; = \; \big( D_{1,  N+1}(\la) \cdot A_{1,  N+1}(\la-i\hbar) - 1 \big) 
\cdot B_{1,  N+1}^{-1}(\la-i\hbar) \;. 
\enq
\begin{cor}
\label{Corolaire Action C sur Vect Propres B}
The action of the operator $C_{1,N+1}(\la)$ on the function $\Psi_{\bs{y}_N, \veps} (\bs{x}_{N+1})$ takes the form
\bem
C_{1,N+1}(\la) \cdot \Psi_{\bs{y}_N, \veps} (\bs{x}_{N+1}) \; = \; \; \big( \la-\veps + \ov{\bs{y}}_N \big)  
 \sul{p=1}{N} (-i)^{N+1} \pl{ \substack{ s=1 \\ \not=p} }{ N } \Big( \f{\la-y_s}{y_p-y_s} \Big) 
\cdot  \Psi_{\bs{y}_N-i\hbar \bs{e}_p, \veps-i\hbar} (\bs{x}_{N+1}) \\
+ \sul{ \substack{ p,r=1  \\ p \not= r} }{N} \pl{ \substack{ s=1  \\ \not= p,r} }{N} (\la-y_s)
\pl{a=r,p}{} \pl{ \substack{b=1  \\ \not= p,r} }{N} \bigg\{ \f{1}{y_a-y_b} \bigg\}
\cdot \f{ \Psi_{\bs{y}_N+i\hbar (\bs{e}_p-\bs{e}_r), \veps-i\hbar} (\bs{x}_{N+1})  }{ (y_p-y_r)(y_r-y_p-i\hbar) } 
\; + \; \mc{W}(\{y_a\}_1^N  ; \la)  \cdot \Psi_{\bs{y}_N , \veps-i\hbar} (\bs{x}_{N+1}) \;, 
\label{ecriture action operateur C sur Psi}
\end{multline}
where $\mc{W}(\{y_a\}_1^N ; \la)$ is given by 
\beq
 \mc{W}(\{y_a\}_1^N ;  \la) \; = \; 
- \sul{p=1}{N} \pl{ \substack{s=1 \\ \not= p} }{N}  \bigg\{ \f{\la-y_s}{y_p-y_s} \bigg\}
\cdot \sul{ \eps= \pm }{ } \pl{\ell=1}{N} \Big[ \f{1}{y_p-y_{\ell} +i \eps \hbar}\Big] \;. 
\label{ecriture forme coefficient W pour action op C}
\enq
\end{cor}

\Proof 

A direct calculation allows one to recasts the action of the operator $C_{1,\dots,N+1}(\la)$ in the form
\eqref{ecriture action operateur C sur Psi} with 
\beq
\mc{W}(\{y_a\}_1^N  ;  \la) \; = \; \sul{p=1}{N}   \f{1}{ \la - y_p -i\hbar} \cdot 
\pl{\substack{s=1 \\ \not= p} }{N} \bigg\{ \f{\la-y_s}{(y_p - y_s + i\hbar)(y_p-y_s)} \bigg\}
\; - \; \pl{\ell=1}{N}  \Big[ \f{1}{\la- y_{\ell} - i \hbar} \Big] \;. 
\label{ecriture form originale fct W}
\enq
In order to relate \eqref{ecriture form originale fct W}  to \eqref{ecriture forme coefficient W pour action op C}, we 
observe that  $\mc{W}(\{y_a\}_1^N  ;  \la)$  is a rational function of $\la$ such that 
$\mc{W}(\{y_a\}_1^N  ;  \la) \; = \; \e{O}(\la^{N-2})$ when $\la \tend \infty$ and with potential poles at 
$\la=y_{\ell} +i\hbar$, $\ell=1,\dots,N$. However, one has that 
\beq
\e{Res} \Big( \mc{W}(\{y_a\}_1^N  ;  \la) \cdot \dd \la , \la= y_{p}+i\hbar \Big) \; = \; 
\pl{ \substack{ \ell=1 \\ \not= p} }{N}  \Big[ \f{1}{y_p- y_{\ell} }\Big] \; - \; 
\pl{ \substack{ \ell=1 \\ \not= p} }{N}  \Big[ \f{1}{y_p- y_{\ell} }\Big] \; = \; 0 \;. 
\enq
It thus follows that $\mc{W}(\{y_a\}_1^N  ;  \la)$ is a polynomial in $\la$ of degree $N-2$. As such it 
can be reconstructed by an interpolation at the points $\la=y_a$, $a=1, \dots, N$. It is readily seen  that one has
\beq
\mc{W}(\{y_a\}_1^N ; y_r) \; = \; \; - \; \pl{\ell=1}{N}  \Big[ \f{1}{y_r- y_{\ell} + i \hbar} \Big] 
\; - \; \pl{\ell=1}{N}  \Big[ \f{1}{y_r - y_{\ell} - i \hbar} \Big]  \;. 
\enq
Hence the expression \eqref{ecriture forme coefficient W pour action op C} for $\mc{W}(\{y_a\}_1^N ; \la)$ follows. \qed





\section{Resolution of the inverse problem}
\label{Section resolution probleme inverse}

In this section, we show how to express the action of certain specific local operators on 
the kernel $\Psi_{\bs{y}_N,\veps}(\bs{x}_{N+1})$ of the SoV transform in terms of an action on its dual variables. 
In their turn, these dual equations show that the local operators of interest are realized  
as linear combinations of shift and multiplication operatots on 
the space $L_{\e{sym}\times -}^{2}\big(\R^N\times \R, \dd \mu(\bs{y}_N)\otimes \dd \veps\big)$ 
where the quantum separation of variables occurs.

\subsection{The position operators}

Having established a specific representation for the system of eigenfunctions of the $B_{1,\dots,N+1}(\la)$
operator, we are in position to compute the action of a certain class of operators on the 
kernel of the SoV transform  $\Psi_{\bs{y}_N,\veps}(\bs{x}_{N+1})$.
This result allows us to deduce the action on $\Psi_{\bs{y}_N,\veps}(\bs{x})$
of a complete system of local operators. This provides a solution to the inverse problem.

\begin{prop}
\label{proposition action operateur Or sur fct propres B}
The operator 
\beq
\mc{O}_r(\la) \; = \; \pl{a=1}{r} \Big\{  \ex{ x_a - x_{N+1} }  \Big\} \; \cdot  \; D_{r+1,N+1}(\la) 
\enq
 has the below action on 
the kernel  $ \Psi_{\bs{y}_N,\veps}(\bs{x}_{N+1})$ of the SoV transform
\beq
\mc{O}_r(\la) \cdot \Psi_{\bs{y}_N,\veps}(\bs{x}_{N+1}) \; = \; 
- \sul{ \substack{  \mc{I}_N =  \sg \cup  \ov{\sg} \\ \#\sg = r+1} }{}  \pl{ \substack{ a\in \sg \\ b\not\in \sg }}{} \bigg\{ \f{- i}{y_a-y_b} \bigg\}
\cdot  \pl{ b\not\in \sg }{} (\la-y_b)  \cdot \Psi_{\bs{y}_N-i\hbar \sul{a \in \sg }{} \bs{e}_a,\veps}(\bs{x}_{N+1}) \;, 
\label{ecriture action operateur O r sur Psi}
\enq
where $\mc{I}_N = \intn{1}{N}$ and the sum runs through all partitions $\sg \cup \ov{\sg}$ of $\mc{I}_N$ under the constraint
$\# \sg = r+1$. 

\end{prop}

We postpone the proof of the above proposition to Appendix \ref{Appendix Proof prop action op Or}.

Note that the formula for the action of the operator $\mc{O}_r(\la)$ given in \eqref{ecriture action operateur O r sur Psi} 
allows one to access to the one for products of position operators. 
Indeed, one has the
\begin{cor}
The operator $\pl{a=1}{r} \Big\{  \ex{ x_a - x_{N+1} }  \Big\} $ has the below action on 
$ \Psi_{\bs{y}_N,\veps}(\bs{x}_{N+1})$
\beq
\pl{a=1}{r} \Big\{  \ex{ x_a - x_{N+1} }  \Big\} \cdot \Psi_{\bs{y}_N,\veps}(\bs{x}_{N+1}) \; = \; 
 \sul{ \substack{ \mc{I}_N=\sg \cup \ov{\sg} \\ \#\sg = r} }{}  \pl{ \substack{ a\in \sg \\ b\not\in \sg }}{} 
 \bigg\{ \f{- i}{y_a-y_b} \bigg\}
\cdot \Psi_{\bs{y}_N-i\hbar \sul{a \in \sg }{} \bs{e}_a,\veps}(\bs{x}_{N+1}) \;. 
\label{ecriture action operateur position 1 vers r sur Psi}
\enq
\end{cor}

Note that \eqref{ecriture action operateur position 1 vers r sur Psi} is precisely the equation in dual variables obtained by 
Babelon in \cite{BabelonActionPositionOpsWhittakerFctions}

\Proof

This is a straightforward consequence  of formula \eqref{ecriture action operateur O r sur Psi}
as soon as one observes that 
\beq
D_{r,N+1}(\la)  \; = \; - \la^{N-r} \ex{x_{r} - x_{N+1}} \; + \; \e{O}\big( \la^{N-r-1} \big)  \; .
\enq

\qed 

\subsection{Reconstruction of the momentum operators}

\begin{prop}
The following action in dual variables holds 
\beq
\pl{\ell=1}{r} \Big\{ \ex{x_{\ell}-x_{N+1}} \Big\} \cdot \Big( \sul{\ell=1}{r} p_{\ell} \Big) \cdot
 \Psi_{\bs{y}_N; \veps}(\bs{x}_{N+1}) \; = \;
 \sul{ \substack{  \mc{I}_N=\sg \cup \ov{\sg} \\ \#\sg = r} }{} \; 
 \pl{ \substack{ a\in \sg \\ b\not\in \sg }}{} \bigg\{ \f{- i}{y_a-y_b} \bigg\}
\, \cdot   \,\Big( \sul{a \in \sg}{} y_a \Big)  \cdot 
\Psi_{\bs{y}_N-i\hbar \sul{a \in \sg }{} \bs{e}_a,\veps}(\bs{x}_{N+1})
\label{ecriture action operateurs impuslion sur fonction propre chaine ouverte}
\enq
\end{prop}

The form of the above action, expressed directly in terms of the action of the operator on 
$L_{\e{sym}\times -}^{2}\big(\R^N\times \R, \dd \mu(\bs{y}_N)\otimes \dd \veps\big)$, 
has been first proposed by Babelon in \cite{BabelonQuantumInverseProblemConjClosedToda}. 
Recently, the formula \eqref{ecriture action operateurs impuslion sur fonction propre chaine ouverte}
has been established by Sklyanin through a slightly different procedure \cite{SklyaninResolutionIPFromQDet}. 

\Proof

A Lax matrix acting on any quantum site $r$ of the full chain can be inverted with the help of the quantum determinant relation
\eqref{formule inverse op de Lax}. 
This relation allows one to obtain a recursive reconstruction for the entries of the monodromy matrix 
$T_{0;r+1,\dots,N+1}(\la)$ in terms of the entries of the Lax matrix on site $r$ and the 
monodromy matrix $T_{0;r,\dots,N+1}(\la)$. Indeed, one has
\beq
T_{0;r+1,N+1}(\la) \; = \; \big[  L_{0r}(\la) \big]^{-1} \cdot  T_{0;r,N+1}(\la)   \;. 
\label{relation entre mat mon r+1 et r}
\enq

The idea of such a recursive reconstruction has been first proposed by Kuznetzov 
\cite{KuznetsovIPforclassicalSL(2)ChainsRecReconstrSepVars} when he considered the inverse problem for
classical sl(2)-type integrable lattices. Relations such as \eqref{relation entre mat mon r+1 et r} 
allowed him to provide a recursive construction of the canonical transformation between the classical separated variables 
for a chain of length $N$ and $N+1$. 

In our case of interest, the above identity allows us to access to the action of the combination of position and momenta of interest
on the function $\Psi_{\bs{y}_N;\eps}(\bs{x}_{N+1})$. Namely, the equality between the second columns in 
\eqref{relation entre mat mon r+1 et r} leads to a three-term recurrence relation satisfied by $D$-operators
associated with chains of different lengths:
\beq
D_{r+1,N+1}(\la)  \; = \; -\ex{x_r-x_{r-1} } D_{r-1,N+1}(\la) \, + \,(\la-i\hbar -p_r) \,  D_{r,N+1}(\la) 
\qquad  \e{for}  \qquad r=2,\dots, N \;, 
\label{ecriture relation de recurrence trois termes ops D}
\enq
and the initiation condition
\beq
D_{2,N+1}(\la)  \; = \; \ex{x_1} B_{1,N+1}(\la) \, + \,(\la-i\hbar -p_1) \,  D_{1,N+1}(\la)  \;. 
\label{ecriture relation D2 D1 etr B1}
\enq

\noindent The operator $D_{r,\dots,N+1}(\la)$ is an operator-valued polynomial in $\la$ of degree $N-r$\, :
\beq
D_{r,N+1}(\la) \; = \; \sul{p=0}{N-r} \la^{N-r-p}  \cdot d_p^{(r)} \;. 
\enq
It is easy to check that 
\beq
\lim_{\la \tend \infty} \Big[  \la^{r-N}D_{r,N+1}(\la) \Big] \; = \;  - \ex{x_r- x_{N+1} }  \qquad \e{so} \; \e{that} 
\qquad 
d_0^{(r)} \; = \; - \ex{x_r- x_{N+1} } \;. 
\enq

The \textit{lhs} of equation \eqref{ecriture relation de recurrence trois termes ops D} 
is an operator-valued polynomial in $\la$ of degree $N-r-1$. Since, \textit{a priori}, the \textit{rhs} of this equation is 
an operator valued polynomial in $\la$ of degree $N-r+1$, its coefficients associated with 
$\la^{N-r+1}$ and $\la^{N-r}$ have to vanish. 
More precisely, \eqref{ecriture relation de recurrence trois termes ops D} implies that  
\bem
\e{O}\big( \la^{N-r-1} \big)  \; = \; \ex{x_r-x_{N+1}} \la^{N-r+1}  \; - \;  \ex{x_r-x_{r-1}} d_1^{(r-1)} \la^{N-r}
 \; - \;   \la^{N-r+1}    \ex{x_r-x_{N+1}}    \\
\; + \;  \la^{N-r} d_1^{(r)} -(i\hbar + p_r) \times  - \ex{x_r-x_{N+1}}  \la^{N-r}  \; + \; \e{O}\big( \la^{N-r-1} \big)  \;. 
\end{multline}
The leading terms cancel out explicitly whereas the cancellation of the first sub-leading ones leads to 
the relation
\beq
\ex{ x_r - x_{N+1} } p_r   \; + \; d_1^{(r)} \; - \; \ex{x_r - x_{r-1} } d_1^{(r-1)} \; = \; 0 
\qquad \quad \e{for} \quad \qquad r=2,\dots, N \;. 
\label{ecriture relation de recurrence pour calculer les p}
\enq
We also need the relation at $r=1$ so as to have a closed system. The operator 
$B_{1, N+1}(\la)$ is a polynomial in $\la$ of degree $N$ with leading asymptotics 
at infinity $\ex{-x_{N+1}} \la^{N}$. It has the representation
\beq
B_{1, N+1}(\la)  \; = \; \ex{-x_{N+1}} \sul{p=0}{N}  \la^{N-p} b_p  \;. 
\enq
Inserting this decomposition into \eqref{ecriture relation D2 D1 etr B1} and using that the 
coefficients in front of $\la^N$ and $\la^{N-1}$ in \eqref{ecriture relation D2 D1 etr B1} both vanish leads to 
\beq
\ex{ x_1 - x_{N+1} }  p_1  \; + \; d_1^{(1)} \; + \; \ex{x_1 - x_{N+1} } b_1 \; = \; 0 \;. 
\label{ecriture relation entre p1 d1 et b1}
\enq
A decreasing induction shows that 
\beq
\pl{\ell=t-1}{r} \Big\{ \ex{x_{\ell}-x_{N+1}}  \Big\} \cdot \Big( \sul{ \ell=t}{ r} p_{\ell} \Big) 
\; + \; \pl{\ell=t-1}{r-1} \Big\{ \ex{x_{\ell}-x_{N+1}}  \Big\} \cdot  d_1^{(r)}
\; - \;  \pl{\ell=t}{r} \Big\{ \ex{x_{\ell}-x_{N+1}}  \Big\} \cdot  d_1^{(t-1)} \; = \; 0
\quad \e{for} \; \; r \geq t \geq 2 \;. 
\label{ecriture relation entre p et DA divers d sur plusieurs sites}
\enq
Hence, setting $t=2$ and replacing $d_1^{(1)}$ with the help of \eqref{ecriture relation entre p1 d1 et b1} gives 
\beq
\pl{\ell=1}{r} \Big\{ \ex{x_{\ell}-x_{N+1}}  \Big\} \cdot \Big( \sul{ \ell=1}{ r} p_{\ell} \Big) 
\; = \; - \pl{\ell=1}{r} \Big\{ \ex{x_{\ell}-x_{N+1}}  \Big\} \cdot  b_1
\; - \;  \pl{\ell=1}{r-1} \Big\{ \ex{x_{\ell}-x_{N+1}}  \Big\} \cdot  d_1^{(r)}   \;. 
\label{ecriture relation entre p et DA divers d sur plusieurs sites}
\enq
Equation \eqref{ecriture relation entre p et DA divers d sur plusieurs sites} above recasts 
the combination of position and momentum operators of interest solely in terms
of operators whose action on the function $\Psi_{\bs{y}_N;\eps}(\bs{x}_{N+1})$ is known. Indeed, the action of the product of
exponents in position operators in given by  \eqref{ecriture action operateur position 1 vers r sur Psi}, whereas the action of 
the coefficients $b_1$ and $d_1^{(r)}$ can be deduced, respectively, from  \eqref{ecriture action B sur Psi} and 
\eqref{ecriture action operateur O r sur Psi} which yields
\beq
b_1 \cdot \Psi_{\bs{y}_N,\veps} (\bs{x}_{N+1}) \; = \; - \ov{\bs{y}}_N \cdot \Psi_{\bs{y}_N,\veps-i\hbar} ( \bs{x}_{N+1} )
\enq
and 
\beq
\pl{\ell=1}{r-1} \Big\{ \ex{x_{\ell}-x_{N+1}}  \Big\} \cdot d_1^{(r)} \cdot \Psi_{\bs{y}_N,\veps}(\bs{x}_{N+1}) \; = \; 
 \sul{ \substack{ \mc{I}_N=\sg  \cup \ov{\sg} \\ \#\sg = r} }{}  
 \pl{ \substack{ a\in \sg \\ b\not\in \sg }}{} \bigg\{ \f{- i}{y_a-y_b} \bigg\}
\cdot  \Big( \sul{b \not\in \sg }{} y_a \Big)  \cdot \Psi_{\bs{y}_N-i\hbar \sul{a \in \sg }{} \bs{e}_a,\veps}(\bs{x}_{N+1}) \;. 
\enq

It only remains to carry out a straightforward replacement. \qed

\subsection{Reconstruction of other operators}

The below lemma has been established recently by Sklyanin \cite{SklyaninResolutionIPFromQDet}. 

\begin{lemme}

Given any $N$ and $r\in \intn{2}{N-1}$ has the below operator identities
\beqa
A_{1,r}(\la) & = & D_{r+1,N+1}(\la+i\hbar) \cdot A_{1,N+1}(\la) \;  - \; C_{r+1,N+1}(\la+i\hbar) \cdot B_{1,N+1}(\la)  
\label{equation definissant A 1 a r} \\
B_{1,r}(\la) & = & A_{r+1,N+1}(\la+i\hbar) \cdot B_{1,N+1}(\la) \;  - \; B_{r+1,N+1}(\la+i\hbar) \cdot A_{1,N+1}(\la)  \\
C_{1,r}(\la) & = & D_{r+1,N+1}(\la+i\hbar) \cdot C_{1,N+1}(\la) \;  - \; C_{r+1,N+1}(\la+i\hbar) \cdot D_{1,N+1}(\la)  
\label{equation definissant C 1 a r}\\
D_{1,r}(\la) & = & A_{r+1,N+1}(\la+i\hbar) \cdot D_{1,N+1}(\la) \;  - \; B_{r+1,N+1}(\la+i\hbar) \cdot C_{1,N+1}(\la)  \; . 
\eeqa

\end{lemme}

\Proof 

We only establish the first identity. The other ones are proved in much the same way. 
One has, due to the quantum determinant relation, 
\bem
A_{1,r}(\la) \; = \; A_{1,r}(\la)  \cdot \Big(  D_{r+1,N+1}(\la+i\hbar)   A_{r+1,N+1}(\la) \; - \; 
 C_{r+1,N+1}(\la+i\hbar)   B_{r+1,N+1}(\la)   \Big)  \\
 \; = \;  D_{r+1,N+1}(\la+i\hbar) \cdot \Big( A_{1,N+1}(\la) - B_{1,r}(\la) C_{r+1,N+1}(\la) \Big)  \; + \; 
 C_{r+1,N+1}(\la+i\hbar) A_{1,r}(\la)  B_{r+1,N+1}(\la)   \\
\; = \; D_{r+1,N+1}(\la+i\hbar) \cdot A_{1,N+1}(\la) \;  - \; C_{r+1,N+1}(\la+i\hbar) \cdot B_{1,N+1}(\la) \; , 
\end{multline}
where, to get the last line, we have used the off-diagonal quantum-determinant issued equations.

\begin{prop}
The operator 
\beq
\msc{A}_r(\la) \; = \; \pl{a=1}{r} \Big\{  \ex{ x_a - x_{N+1} }  \Big\} \; \cdot  \; A_{1,r}(\la) 
\enq
 has the below action on 
the kernel  $ \Psi_{\bs{y}_N,\veps}(\bs{x}_{N+1})$ of the SoV transform
\beq
\msc{A}_r(\la) \cdot \Psi_{\bs{y}_N,\veps}(\bs{x}_{N+1}) \; = \; 
 \sul{ \substack{  \mc{I}_N =  \sg \cup  \ov{\sg} \\ \#\sg = r} }{}  
\pl{ \substack{ a\in \sg \\ b\not\in \sg }}{} \bigg\{ \f{- i}{y_a-y_b} \bigg\}
\cdot  \pl{ b\in \sg }{} (\la-y_b)  \cdot \Psi_{\bs{y}_N-i\hbar \sul{a \in \sg }{} \bs{e}_a,\veps}(\bs{x}_{N+1}) \;, 
\label{ecriture action operateur Ar cal sur Psi}
\enq
where $\mc{I}_N = \intn{1}{N}$. 

\end{prop}

Note that equation \eqref{ecriture action operateurs impuslion sur fonction propre chaine ouverte} can be readily
deduced from the above action of $\msc{A}_r(\la)$ by identifying the $\e{O}(\la^{r-1})$ part of both side's
$\la \tend \infty$ asymptotics.

\Proof

In order to recast $\msc{A}_r(\la) \cdot \Psi_{\bs{y}_N,\veps}(\bs{x}_{N+1})$ in terms of an action on dual variables, 
we use that it is a polynomial in $\la$ of degree $r$. It can thus be reconstructed by interpolation at $r+1$ points. 
For $r=N$ one has
\beq
\msc{A}_N(\la) \; = \; \pl{a=1}{N} \Big\{  \ex{ x_a - x_{N+1} }  \Big\} \cdot \ex{x_{N+1}} \cdot B_{1,N+1}(\la) \;, 
\enq
and the action of all the operators is known. Hence, it remains to consider the case $r<N$. 
In such a case, one obtains $N$ interpolation points $y_1,\dots, y_N$ by  acting  on $\Psi_{\bs{y}_N; \veps}(\bs{x}_{N+1}) $ with
the operator 
\beq
\msc{A}_r(y_a) \; = \; \mc{O}_r(y_a+i\hbar) A_{1,N+1}(y_a)\;   - \; 
\pl{a=1}{N} \Big\{  \ex{ x_a - x_{N+1} }  \Big\} \, \cdot \,  C_{r+1,N+1}(y_a+i\hbar) \cdot B_{1,N+1}(y_a)  
\enq
and using that the second operator produces vanishing contributions. This leads to 
\beq
\msc{A}_r(y_a) \cdot \Psi_{\bs{y}_N,\veps}(\bs{x}_{N+1}) \; = \; 
(-1)^{r} \sul{ \substack{  \mc{I}_N =  \sg \cup  \ov{\sg} \\ \#\sg = r  \\ \ov{\sg} \; / \; a \in \ov{\sg} } }{}  
\pl{ \substack{ s\in \sg \\ t\not\in \sg }}{} \bigg\{ \f{- i}{y_s-y_t} \bigg\}
\cdot  \pl{ s\in \sg }{} (y_s-y_a)  \cdot \Psi_{\bs{y}_N-i\hbar \sul{s \in \sg }{} \bs{e}_s,\veps}(\bs{x}_{N+1}) \;. 
\enq
Hence, reconstructing the polynomial of interest through interpolation, we get 
\beq
\msc{A}_r(\la) \cdot \Psi_{\bs{y}_N,\veps}(\bs{x}_{N+1}) \; = \; 
 \sul{ \substack{  \mc{I}_N =  \sg \cup  \ov{\sg} \\ \#\sg = r} }{}  
\pl{ \substack{ s\in \sg \\ t\not\in \sg }}{} \bigg\{ \f{- i}{y_s-y_t} \bigg\}
\cdot \pl{p \in \sg }{} (\la-y_p) \cdot P_{ \{\ov{\sg}\} }(\la) 
 \cdot \Psi_{\bs{y}_N-i\hbar \sul{s \in \sg }{} \bs{e}_s,\veps}(\bs{x}_{N+1}) \;, 
\enq
where 
\beq
P_{ \{\ov{\sg}\} }(\la) \; = \; \sul{ a \in \ov{\sg} }{}  
\pl{ \substack{ p \in \ov{\sg} \\ p \not= a} }{} \bigg\{ \f{\la - y_p }{ y_a - y_p } \bigg\} \;. 
\enq
Again, by interpolation, it is readily seen that $P_{ \{\ov{\sg}\} }(\la)=1$. \qed

\begin{cor}
\bem
 \pl{a=1}{r} \Big\{  \ex{ x_a - x_{N+1} }  \Big\} C_{r+1,N+1}(\la) \cdot \Psi_{\bs{y}_N,\veps}(\bs{x}_{N+1}) \; = \; 
- \big( \la - \veps  + \ov{\bs{y}}_N \big) \sul{ \substack{  \mc{I}_N =  \sg \cup  \ov{\sg} \\ \#\sg = r+1} }{}  
\pl{ \substack{ a\in \sg \\ b \in \ov{\sg} }}{} \bigg\{ \f{- i}{y_a-y_b} \bigg\}
\cdot  \pl{ b\in \ov{\sg} }{} (\la-y_b)  \cdot \Psi_{\bs{y}_N-i\hbar \bs{e}_{\sg} ,\veps-i\hbar}(\bs{x}_{N+1}) \\
\; + \; \sul{ \substack{  \mc{I}_N =  \sg \cup  \ov{\sg} \\ \#\sg = r} }{}  
\pl{ \substack{ a\in \sg \\ b\in \ov{\sg} }}{} \bigg\{ \f{- i}{y_a-y_b} \bigg\}
\cdot  \mc{W}\big( \{y_a\}_{ a \in \ov{\sg}} \; ; \; \la  \big) 
\cdot \Psi_{\bs{y}_N-i\hbar \bs{e}_{\sg} ,\veps-i\hbar}(\bs{x}_{N+1}) \\
\sul{ \substack{  \mc{I}_N =  \sg \cup  \ov{\sg} \cup \{ p \}  \\ \#\sg = r+1} }{}  
\pl{ \substack{ a\in \sg \\ b \in \ov{\sg} }}{} \bigg\{ \f{- i}{y_a-y_b} \bigg\}
\pl{a \in \sg }{} \bigg\{ \f{- i}{y_a - y_p - i\hbar } \bigg\}
\pl{a \in \sg \cup \ov{\sg}  }{} \bigg\{ \f{- i}{y_a - y_p  } \bigg\}
\cdot  \pl{ b\in \ov{\sg} }{} (\la-y_b) 
 \cdot \Psi_{\bs{y}_N +i\hbar (\bs{e}_p- \bs{e}_{\sg})  ,\veps-i\hbar}(\bs{x}_{N+1}) 
\end{multline}
in which $\mc{W}$ is as given by \eqref{ecriture forme coefficient W pour action op C}.  Also, 
we do stress that the last sum over partitions runs with the integer $p$ not being fixed. 
Finally, we have introduced the notation
\beq
\bs{e}_{\sg} \; \equiv \; \sul{a \in \sg }{}  \bs{e}_{ a } \;.   
\enq

\end{cor}

This formula is a mere conjunction of \eqref{ecriture action operateur Ar cal sur Psi} 
and \eqref{ecriture action operateur O r sur Psi} and the relation \eqref{equation definissant A 1 a r}
followed by an application of the two representation \eqref{ecriture form originale fct W} and 
\eqref{ecriture forme coefficient W pour action op C} for $\mc{W}\big( \{y_a \}_1^N; \la \big)$. 
Note that a similar formula can be derived for the action of the operator 
\beq
 \pl{a=1}{r} \Big\{  \ex{ x_a - x_{N+1} }  \Big\} C_{1,r}(\la)
\enq
by using equation \eqref{equation definissant C 1 a r} along with the aforeobtained results. 



\section*{Conclusion}

In this paper we have recast the action of various local and non-local operators of the closed Toda
on the SoV transform's kernel $\Psi_{\bs{y}_N;\veps}(\bs{x}_{N+1})$ in terms of an action on the dual variables 
$\bs{y}_N$ and $\veps$. 
Our approach builds on Babelon's idea relative to acting with products of positions operators 
on the integral kernel of a fully iterated Mellin-Barnes multiple integral representation for 
$\Psi_{\bs{y}_N;\veps}(\bs{x}_{N+1})$ \cite{BabelonActionPositionOpsWhittakerFctions}. 
In the present paper, we have managed to make the method more efficient by using $r$-split Mellin-Barnes
representations that are adapted to the type of operators for which one wants to set an
equation in dual variables. 
The Mellin-Barnes based approach appears to be quite systematic and 
the obtained formulae take a quite universal form. It is thus plausible that the present results should 
extend quite easily to the case of other quantum integrable models solvable by the quantum separation of variables
method. Furthermore, when combined with Sklyanin's quantum determinant based identities, it allows one to obtain 
equations in dual variables for an even larger family of operators associated with the model.

\section*{Acknowledgements}

K. K. K. would like to thank E. K. Sklyanin  
for stimulating discussions relative to various topics treated in this paper. 
K. K. K. is supported by CNRS, the Burgundy region PARI 2013 FABER grant  
"Structures et asymptotiques d'int\'{e}grales multiples" and the PEPS-PTI 2012"Asymptotique d'int\'{e}grales multiples" grant. 
This research has been initiated when the author was supported by the EU Marie-Curie Excellence Grant MEXT-CT-2006-042695 and DESY.
K. K. K. is indebted to the Laboratoire de Physique Th\'{e}orique d'Annecy-Le-Vieux for its warm hospitality
when part of the research has been carried out.

\appendix 

\section{Proof of Proposition \ref{Proposition Estimates for decay at infinity of eigenfunction B}}
\label{Appendix Proof prop decay eigenfcts B op}

\Proof

We prove the claim by induction. Since the independence of the construction on $r$ will follow from the next proposition,
we take this fact for granted here. 
It is clear that all the claims of the proposition are satisfied for $N=0$. 
Assume that they hold true up to some $N$, this for any $r \in \intn{1}{N-1}$. It is well know that there exists $C_{\eps,\eta}>0$ such that one has the bounds 
\beq
\big| \Ga(x+iy) \big|   \; \leq  \; C_{\eps,\eta} \cdot |y|^{x-\f{1}{2}} \ex{-\f{\pi}{2} |y| } \qquad
\e{uniformly} \; \e{in} \; y \in \R \;, \quad \eta > |x| > \eps > 0 \; .  
\enq
Likewise, there also exists  $C_{\eta}>0$ such that 
\beq
\big| \Ga(x+iy) \big|^{-1}   \; \leq  \; C_{\eta} \cdot |y|^{\f{1}{2}-x} \ex{\f{\pi}{2} |y| } \qquad
\e{uniformly} \; \e{in} \; y \in \R \; \; \e{and} \;\; |x|< 2\eta \;. 
\enq

Thus we pick an $\eta>0$ such that $\eta \, > \,  \max_{k \in \intn{1}{N}} \big( |\Im(y_k)| \big)  $, and we fix the
shift $\a$ occurring in \eqref{definition du contour d'intégration pour fcton Whittaker}. It is then readily seen
that there exists $\eta$-dependent constants $C>0$ and $M, M^{\prime}$ such that 
\bem
|\varpi(\bs{w},\bs{z} \mid \bs{y}_N)| \; \leq  \; C  \big(  1 + \norm{\bs{w}} + \norm{\bs{z}} \big)^M  \cdot 
 \big(  1 + \norm{\bs{y}_N} \big)^{M^{\prime}}
\pl{b=1}{N} \bigg\{ \ex{ -\f{\pi}{2\hbar} \big( \sul{a=1}{r-1} |\Re(w_a-y_b)| \; + \; \sul{a=r}{N-1} |\Re(z_a-y_b)| \big) } \bigg\}
\\
\times \pl{a=1}{r-1}\pl{b=r}{N-1} \bigg\{  \ex{ \f{\pi}{2\hbar}  |\Re(z_b-w_a)|      }  \bigg\} \cdot 
\ex{ \f{\pi}{2\hbar} \big( \sul{a,b=1}{r-1} |\Re(w_a-w_b)| \; + \; \sul{a,b=r}{N-1} |\Re(z_a-z_b)| \big) } \cdot 
\end{multline}
Note that above and in the following, the dimensionality of the $L^{1}$-norms , \textit{cf}. \eqref{definition normes L1 vecteurs},
is undercurrent by the context. 
Since the induction hypothesis does not depend on $r$, one does not have to discuss which pattern of decomposition has been 
used so as to construct the functions appearing in the integral representation \eqref{ecriture rep int fct propre B}. 
Thus choosing some $r$ and applying the induction hypothesis to these two functions, one gets that there exist constants 
$\kappa_1, \kappa_2,  \kappa_3$ and $\kappa_4$ such that 
\beq
\big| \Psi_{ \bs{w} ; \ov{\bs{y}}_N-\ov{\bs{z}} }(\bs{x}_1) \Psi_{\bs{z} ; \veps-\ov{\bs{y}}_N+\ov{\bs{z}} }(\bs{x}_2) \;  
\varpi(\bs{w},  \bs{z} \mid \bs{y}_N)  \big|  \; \leq \; 
 \kappa_1 \ex{\kappa_2 \norm{x_{N+1}}}  \cdot  \big(  1 + \norm{\bs{y}_N} \big)^{\kappa_4}
  \big(  1 + \norm{\bs{\ga}} \big)^{\kappa_3}   
\cdot \ex{  -\f{\pi}{2 \hbar}  \Big( \sul{b=1}{N} \sul{a=1}{N-1} |\la_b-\ga_a|  
\, - \, \sul{ a<b }{N-1} | \ga_a - \ga_b |   \Big)   } \;.  
\label{ecriture bound on product whittaker functions}
\enq
Above, we have reparameterized the variables $\bs{y}_N$, $\bs{w}$ and $\bs{z}$
\beq
\la_b \; = \; \Re(y_b) \quad \e{and} \quad \bs{\ga} \; = \; \big(\Re(w_1), \dots ,\Re(w_{r-1}), \Re(z_r),\dots , \Re(z_{N-1}) \big)\;, 
\enq
Also, the eventual dependence on $r$ of the constants $\kappa_a$, $a=1,\dots,4$ was removed by taking the supremum
when $r$ runs through $\intn{1}{N-1}$.

Since the upper bound above is already symmetric in respect to the $\la_a's$ and $\ga_a$'s, 
it is enough to show that it belongs to $L^1\big( \Om_{N-1} \big)$ where 
\beq
\Om_{N-1} \; = \; \big\{ \bs{\ga}_{N-1} \in \R^{N-1} \; : \;  \ga_1 < \dots <\ga_{N-1}   \big\} \;. 
\enq
Further, again in virtue of the symmetry, in doing so, one may 
also  assume the ordering  $\la_1 < \dots  < \la_N$. For such an ordering of both sets of 
variables, one has the identity \cite{IorgovShaduraIntRepEigenFctonsBopTodaWithBdry}
\beq
\sul{k=1}{N}\sul{j=1}{N-1} |\ga_j - \la_k| \; - \; \sul{a<b}{N}|\la_a - \la_b|  \; - \; \sul{a<b}{N-1}|\ga_a - \ga_b|  \; = \;
\sul{j=1}{N-1} \phi_j(\ga_j \mid \bs{\la}_N) 
\enq
where 
\beq
\phi_j(\ga_j \mid \bs{\la}_N ) \; = \;  \sul{k=1}{j} \Big[|\ga_j-\la_k| -(\la_k-\ga_j)   \Big] \; + \;  
\sul{k=j+1}{N} \Big[|\ga_j-\la_k| - (\ga_j-\la_k) \Big]  \;. 
\enq
Note that this identity, when the \textit{rhs} has been 
replaced with $\geq 0$, has been first used in \cite{GerasimovKharchevLebedevRepThandQISM}
so as to prove the convergence properties of the integral in question.
Its use leads to 
\beq
\big| \Psi_{ \bs{w} ; \ov{\bs{y}}_N-\ov{\bs{z}} }(\bs{x}_1) \Psi_{\bs{z} ; \veps-\ov{\bs{y}}_N+\ov{\bs{z}} }(\bs{x}_2) \;  
\varpi(\bs{w},  \bs{z} \mid \bs{y}_N)  \big|  \; \leq \; 
 \kappa_1 \ex{\kappa_2 \norm{x_{N+1}}}  \cdot  \big(  1 + \norm{\bs{y}} \big)^{\kappa_4}
\pl{a<b}{N} \ex{  -\f{\pi}{2 \hbar}   |\Re(y_a-y_b)| } \cdot  \pl{j=1}{N-1}  
\bigg\{   \big(  1 + |\ga_j | \big)^{\kappa_3}  \ex{  -\f{\pi}{2 \hbar} \phi_j(\ga_j \mid \bs{\la}_N ) }
\bigg\}
\label{ecriture majorant global de l'integrande pour mellin barnes des Whittaker}
\enq
for $\bs{\ga}_{N-1} \in \Om_{N-1}$. Hence, we consider the integral 
\beq
\msc{I}_{N}   \; = \; \Int{ \R^{N-1} }{}  \big(  1 + \norm{\bs{\ga}} \big)^{\kappa_3}  
\cdot  \pl{j=1}{N} \exp\Big\{  -\f{\pi}{2 \hbar} \phi_j(\ga_j \mid \bs{\la}_N ) \Big\}  \cdot \bs{1}_{\Om_{N-1}}(\bs{\ga}_{N-1})
 \cdot  \dd^N \ga  \;. 
\label{ecriture integrale multiple majoree pour bonne definition fct propre de B}
\enq
and compute the integrals successively starting from the one over $\ga_{N-1}$. Since the integrand is positive and
piecewise continuous on $\R^{N-1}$ and Lebesgue's  measure is $\sg$-finite, by Fubbini-Tonelli-Lebesgue theorem 
this is enough so as to guarantee the $L^1(\R^{N+1})$ character of the integrand. For this purpose, given any $j$
and $m_j \in \R$, observe that one has the chain of majorations:
\bem
\Int{ \ga_{j-1} }{ + \infty } \big( 1+|\ga_j | \big)^{m_j} \cdot 
\ex{ - \f{\pi }{2 \hbar  } \phi_j(\ga_j \mid \bs{\la} )  }  \cdot \dd \ga_j
\; \leq  \; \Int{ \min(\ga_{j-1}, \la_1 )  }{  \max(\ga_{j-1}, \la_N )  } \big( 1+|\ga_j | \big)^{m_j} \cdot \dd \ga_j 
\; + \;  \; \Int{  \max(\ga_{j-1}, \la_N ) }{+\infty}  \big( 1+|\ga_j | \big)^{m_j} 
  \cdot \ex{ - \f{\pi }{2 \hbar  } \big(2j \ga_j - 2 \sul{k=1}{j} \la_k   \big)  }  \cdot \dd \ga_j  \\
\leq \; \f{1}{m_j+1} \Big[ \big(1+  \max(\ga_{j-1}, \la_N ) \big)^{m_j+1} \; - \; \big(1+  \min(\ga_{j-1}, \la_1 ) \big)^{m_j+1}  \Big] 
\; + \; C_{m_j,N}^{(j)}  \cdot \pl{k=1}{j} \ex{\f{\pi \la_j}{\hbar} }  \; \leq \; 
\wt{C}_{m_j,N}^{(j)} \big(1+  |\ga_{j-1}| ) \big)^{m_j+1}  \; , 
\label{ecriture chaine majoration pour integral sur variable gamma j}
\end{multline}
for some constants $C_{m_j,N}^{(j)}, \wt{C}_{m_j,N}^{(j)}>0$. 
Hence, when carrying out the succesive chain of integrations in 
\eqref{ecriture integrale multiple majoree pour bonne definition fct propre de B} and dealing with the 
integral in respect  to $\ga_j$, it is readily seen that the sole effect of integrating in respect to 
$\ga_N, \dots, \ga_{j+1}$ was to increase the original exponent $\kappa_3$ by some sufficiently large integer, 
thus, effectively, reducing the integration versus $\ga_j$ to the model integral that was written in the \textit{rhs} of 
\eqref{ecriture chaine majoration pour integral sur variable gamma j}.

One can continue in such a way up to integrating over $\ga_1$. Then one deals with an integration over $\R$. 
It is then easy to see that there exists a $\bs{\la}_N$-dependent constant $\wt{C}$ such that 
\beq
\phi_1(\ga_1\mid  \bs{\la}_N ) \;  \geq  \; (2 |\ga_1 | -  \wt{C} ) \;.  
\enq
This last estimate ensures that the integral over $\ga_1$ is convergent as well. 

It solely remains to establish that 
$\Psi_{\bs{y}_N,y_{N+1}}(\bs{x}_{N+1})$ is an entire function of $(\bs{x}_{N+1}, \bs{y}_N,y_{N+1})$. 
This is clear for $N=0$. Further, assume that this has been established up to some $N-1$.
The estimates for the convergence of the integral readily lead to the fact that it defines 
a continuous function of $(\bs{x}_{N+1}, \bs{y}_N,y_{N+1})$.

Let $\msc{C}$ be a closed loop in $\Cx$ lying in the strip $|\Im(z)| < \eta$. Then, since $\msc{C}$
is compact and the recursive integrand converges uniformly in $(\bs{y}_N,y_{N+1})$ and $\bs{x}_{N+1}$
bounded, one gets that, for any $a=1,\dots, N+1 $
\beq
\Oint{ \msc{C} }{} \Psi_{\bs{y}_N,y_{N+1} }(\bs{x}_{N+1}) \cdot \dd y_a  \; = \;
\Int{ \msc{C}_{r-1; N-r} }{} \bigg\{  \Oint{ \msc{C} }{}\Psi_{ \bs{w} ; \ov{\bs{y}}_N-\ov{\bs{z}} }(\bs{x}_1) 
\Psi_{\bs{z} ; y_{N+1} -\ov{\bs{y}}_N+\ov{\bs{z}} }(\bs{x}_2)   \;  
\varpi(\bs{w},  \bs{z} \mid \bs{y}_N)  \dd y_a  \bigg\} \cdot \pl{a=1}{r-1}\dd w_a  \cdot \pl{a=r}{N-1}\dd z_a   \; = \; 0 \;,
\nonumber
\enq
and likewise
\beq
\Oint{ \msc{C} }{} \Psi_{\bs{y}_N,y_{N+1} }(\bs{x}_{N+1}) \dd x_a  \; = \;
\Int{ \msc{C}_{r-1; N-r} }{} \bigg\{  \Oint{ \msc{C} }{}\Psi_{ \bs{w} ; \ov{\bs{y}}_N-\ov{\bs{z}} }(\bs{x}_1) 
\Psi_{\bs{z} ; y_{N+1} -\ov{\bs{y}}_N+\ov{\bs{z}} }(\bs{x}_2)   \;  
\varpi(\bs{w},  \bs{z} \mid \bs{y}_N)  \dd x_a  \bigg\} \cdot \pl{a=1}{r-1}\dd w_a  \cdot \pl{a=r}{N-1}\dd z_a   \; = \; 0 \;. 
\nonumber
\enq
Hence, by Morera's theorem the function is holomorphic in each of the variables taken singly. 
Thus, by Hartog's theorem, it is a holomorphic function of $(\bs{y}_N,y_{N+1})$
belonging to the poly-strip $ |\Im(y_a)| < \eta$ , $a=1,\dots, N+1$ and 
$ |\Im(x_a)| < \eta$ , $a=1,\dots, N+1$. Since $\eta$
was arbitrary, the claim follows. 
 \qed





\section{Proof of Proposition \ref{propositon action operateurs ABD}}
\label{Appendix Proof Prop action ABD}

It is clear from the explicit definition of $\Psi_{\emptyset, \veps}(x)$ and the expressions for the one-site operators
that the equations \eqref{ecriture action B sur Psi}-\eqref{ecriture action A sur Psi} and the two relations given in 
\eqref{ecriture proprietes action ex x sur fct propre B} are indeed satisfied. 
We now prove the statement by induction. Thus we assume having built all of the lower number of variables functions 
$\Psi_{\bs{y}_k;\veps}\big( \bs{x}_{k+1} \big)$, $k=0,\dots,N-1$ which 
\begin{itemize}
\item are independent of the splitting $r$ used for their construction;
\item satisfy the appropriate analogues of the relations
\eqref{ecriture action B sur Psi}-\eqref{ecriture action A sur Psi} 
and \eqref{ecriture proprietes action ex x sur fct propre B}. 
\end{itemize}
We shall now proceed in two steps. First, we shall establish the form of the action of operators on $ \Psi_{\bs{y}_N,\veps}(\bs{x}_{N+1})$
and then we shall prove the independence of the construction on $r$.

\subsection*{$\bullet$ Action of the operators}

We now check that the action of the $N+1$-site operators $B, D$ and $A$ takes the desired form 
\eqref{ecriture action B sur Psi}-\eqref{ecriture action A sur Psi} and that the two relations given in 
\eqref{ecriture proprietes action ex x sur fct propre B} hold. 
For this, we split the matrix product defining
the N+1 site monodromy matrix into  a product of the monodromy matrices associated with subchains of sites $1,\dots,r$  and 
$r+1,\dots,N+1$ respectively:
\beq
T_{0;1,N+1}(\la) \; = \; T_{0;1,r}(\la) \cdot T_{0;r+1, N+1}(\la)
\label{ecriture decomposition deux sites matrice de monodromie}
\enq

We first start by computing the action of the $B_{1,\dots, N+1}( \la )$ operator. 
It follows from the explicit representation for the local Lax matrices \eqref{ecriture matrice de Lax Toda} that 
$B_{1,N+1}(\la)$ is given by a finite linear combination of at most first order differential operators in 
each of the $x_k$'s. Since, according to Proposition \ref{Proposition Estimates for decay at infinity of eigenfunction B}
$\Psi_{\bs{y}_N;\veps}(\bs{x}_{N+1})$ is holomorphic in respect to $x_k$, one can represent $\Dp{x_k}$
in terms of a compactly supported contour integral operator. Then, the absolute convergence of the integral in 
\eqref{ecriture rep int fct propre B} allows one to apply Fubbini's theorem and exchange the orders of integration.  
In other words it is allowed to exchange the integration in \eqref{ecriture rep int fct propre B} with differentiations 
in respect to any $x_k$. As a consequence, one can move the operator $B_{1,N+1}(\la)$
under the integral sign when computing its action. This last step allows one to use the split-like representation for this
operator given in \eqref{ecriture decomposition deux sites matrice de monodromie} 
along with the formulae for the action of the appropriate operators on lower rank functions so as 
to compute the effect of the action. This produces sums involving various combinations 
of functions  $\Psi_{\bs{z} -i\hbar \bs{e}_p ; \veps-\ov{\bs{y}}_N+\ov{\bs{z}} +i\hbar }(\bs{x}_2) $
or $\Psi_{ \bs{w} + i \hbar \bs{e}_p ; \ov{\bs{y}}_N-\ov{\bs{z}} }(\bs{x}_1) $ with shifts in their 
dual variables $\bs{w}$ or $\bs{z}$. One can then split the resulting integral into several sums since all of the
individual terms converge absolutely 
in virtue of the bounds established in Proposition \ref{Proposition Estimates for decay at infinity of eigenfunction B}. 
Then, one can shift the various integration contours by $\pm i\hbar$ so as to recover, in all integrals,  
the product of functions $\Psi_{ \bs{w} ; \ov{\bs{y}}_N-\ov{\bs{z}} }(\bs{x}_1) 
\cdot \Psi_{\bs{z} ; \veps-\ov{\bs{y}}_N+\ov{\bs{z}} +i\hbar}(\bs{x}_2) $
in each term under the integration sign. Note that the apparent poles at $z_p=z_a$ for 
$a\in \{r,\dots, N-1 \} \setminus \{p\}$ and $w_p= w_a$ for $a \in \{1,\dots,r-1\}\setminus \{ p\}$ which arise in the
intermediate calculations are, in fact, cancelled 
out by the zeroes of the weight function $\varpi(\bs{w},  \bs{z} \mid \bs{y}_N)$. All in all, 
these shifts of contours lead to the integral representation
\beq
B_{1, N+1}(\la)  \cdot \Psi_{\bs{y}_N;\eps}(\bs{x}_{N+1}) \; = \; \Int{ \msc{C}_{r-1,N-r}}{} 
g_{\la}(\bs{w},  \bs{z} \mid \bs{y}_N)
\Psi_{ \bs{w} ; \ov{\bs{y}}_N-\ov{\bs{z}} }(\bs{x}_1) \Psi_{\bs{z} ; \veps-\ov{\bs{y}}_N+\ov{\bs{z}} +i\hbar }(\bs{x}_2)
\;  \cdot \pl{a=1}{r-1}\dd w_a  \cdot \pl{a=r}{N-1}\dd z_a 
\enq
where the function $g_{\la}(\bs{w},  \bs{z} \mid \bs{y}_N)$ is given by
\bem
g_{\la}(\bs{w},  \bs{z} \mid \bs{y}_N) \; = \; \big( \la  -  \ov{\bs{y}}_N + \ov{\bs{z}} + \ov{\bs{w}} \big) \pl{a=1}{r-1}(\la-w_a) 
\cdot \pl{a=r}{N-1}(\la-z_a) \cdot 
\varpi(\bs{w},  \bs{z} \mid \bs{y}_N)  \\
\hspace{2cm}  \;  + \; \sul{p=1}{r-1} (i)^r \pl{ \substack{a=1 \\ \not= p} }{r-1}  \Big( \f{ \la-w_a }{ w_p - w_a -i\hbar} \Big) 
\pl{a=r}{N-1}(\la-z_a)  \cdot \varpi(\bs{w} -i\hbar \bs{e}_p ,  \bs{z} \mid \bs{y}_N)  \\
\; + \; \pl{a=1}{r-1}(\la-w_a) \sul{p=r}{N-1} (-i)^{N-r+1} 
\pl{ \substack{a=r \\ \not= p} }{N-1}  \Big( \f{ \la - z_a }{ z_p - z_a  + i\hbar } \Big) 
\cdot \varpi(\bs{w} ,  \bs{z} +i\hbar \bs{e}_p  \mid \bs{y}_N) \;. 
\label{ecriture fonction g lambda}
\end{multline}

The action of the operator $B_{1,  N+1}(\la)$ will take the form given in \eqref{ecriture action B sur Psi}
as soon as we establish that 
\beq
r_{\la}^{(1)}(\bs{w},  \bs{z} \mid \bs{y}_N) \; = \; 
  g_{\la}(\bs{w},  \bs{z} \mid \bs{y}_N) \; - \; \pl{a=1}{N} ( \la - y_a)  \; \cdot \; \varpi(\bs{w} ,  \bs{z}  \mid \bs{y}_N)  
\label{ecriture representation voulue pour fcton G lambda action B vect Whitt}
\enq
vanishes. Since $r_{\la}^{(1)}(\bs{w},  \bs{z} \mid \bs{y}_N) $ is a polynomial in $\la$ of degree at most $N$, it is enough to show that 
it vanishes at $N+1$ points. Hence, we interpolate at 
$\la= w_{\ell}$, $\ell=1,\dots,r-1$ and  $\la=z_a$, $a=r,\dots,N-1$ and then showing that, 
it has large $\la$ asymptotics $r_{\la}^{(1)}(\bs{w},  \bs{z} \mid \bs{y}_N) \; = \; \e{O}(\la^{N-2})$. 
It follows from the system of equations satisfied by the weight factor 
$\varpi(\bs{w} ,  \bs{z}  \mid \bs{y}_N)$:
\beq
\pl{a=1}{N}(w_{\ell} - y_a) \cdot \varpi(\bs{w} ,  \bs{z}  \mid \bs{y}_N)   = 
(i)^r \pl{ \substack{a=1 \\ \not= \ell} }{r-1}  \Big( \f{ w_{\ell} -w_a }{ w_{\ell} - w_a -i\hbar} \Big) 
\; \cdot \;  \pl{a=r}{N-1}(w_{\ell}-z_a) \; \cdot \;  \varpi(\bs{w} -i\hbar \bs{e}_{\ell} ,  \bs{z} \mid \bs{y}_N) 
\enq
for $\ell=1,\dots,r-1 $ and 
\beq
\pl{a=1}{N}(z_{\ell} - y_a) \cdot \varpi(\bs{w} ,  \bs{z}  \mid \bs{y}_N)  =   (-i)^{N-r+1}  
\pl{a=1}{r-1}(z_{\ell} - w_a) 
\pl{ \substack{a=r \\ \not= \ell } }{N-1}  \Big( \f{ z_{\ell} - z_a }{ z_{\ell} - z_a  + i\hbar } \Big) \cdot 
\varpi(\bs{w} ,  \bs{z} +i\hbar \bs{e}_{\ell}  \mid \bs{y}_N)  
\enq
for $\ell=r,\dots,N-1 \nonumber$, that $r_{\la}^{(1)}(\bs{w},  \bs{z} \mid \bs{y}_N) $ vanishes at the aforementioned interpolation points.

Further, one has that 
\beq
\pl{a=1}{N} (\la - y_a) \, \cdot \, \varpi(\bs{w} ,  \bs{z} \mid \bs{y}_N)  \; = \; 
\Big( \la^N \; - \;  \ov{\bs{y}}_N \cdot \la^{N-1} \; + \; \e{O}(\la^{N-2})  \Big) \, \cdot \, \varpi(\bs{w} ,  \bs{z} \mid \bs{y}_N)  \;. 
\enq
It is also readily seen that the sums occurring in the second and third lines of equation
\eqref{ecriture fonction g lambda} are a $\e{O}(\la^{N-2})$. Hence, the leading and first sub-leading 
terms at $\la\tend \infty$ issue from the first line of 
\eqref{ecriture fonction g lambda} and thus read 
\beq
g_{\la}(\bs{w} ,  \bs{z} \mid \bs{y}_N) \; = \; 
\Big( \la^N \; - \;  \ov{\bs{y}}_N \cdot  \la^{N-1} \; + \; \e{O}(\la^{N-2})  \Big)  \cdot \varpi(\bs{w} ,  \bs{z} \mid \bs{y}_N)\; .
\enq
Accordingly, we get that, indeed $r_{\la}^{(1)}(\bs{w},  \bs{z} \mid \bs{y}_N) \; = \; \e{O}(\la^{N-2})$, 
so that, all in all $r_{\la}^{(1)}(\bs{w},  \bs{z} \mid \bs{y}_N) \; = \; 0$.

We now check that the action of the operator $D_{1, N+1}(\la)$ takes the form \eqref{ecriture action D sur Psi}. 
For this purpose we use the representation for this operator in terms of operators associated with various sub-chains of the model 
that follows from \eqref{ecriture decomposition deux sites matrice de monodromie} and 
proceeding exactly as in the case of the action of the operator $B_{1,N+1}(\la)$ so as to compute
the effect of its action by acting with the appropriate lower number of sites operators
on the functions $\Psi_{ \bs{w}  ; \ov{\bs{y}}_N-\ov{\bs{z}} -i\hbar}(\bs{x}_1)$ or 
$\Psi_{\bs{z} ; \veps-\ov{\bs{y}}_N+\ov{\bs{z}} +i\hbar }(\bs{x}_2) $ arising under the integral sign in 
\eqref{ecriture rep int fct propre B}. Then we reorganize the expression by splitting the 
integrand and shifting the contours in appropriate expressions so as to solely integrate 
the product of functions 
$\Psi_{ \bs{w}  ; \ov{\bs{y}}_N-\ov{\bs{z}} -i\hbar}(\bs{x}_1)
\Psi_{\bs{z} ; \veps-\ov{\bs{y}}_N+\ov{\bs{z}} +i\hbar }(\bs{x}_2) $ at the very end of this procedure.  
It then remains to use the recurrence equations under shifts of its variables satisfied by the weight
factor $\varpi(\bs{w}  ,  \bs{z} \mid \bs{y}_N)$ so as to get that 
\beq
D_{1, N+1}(\la)  \cdot \Psi_{\bs{y}_N;\eps}(\bs{x}_{N+1}) \; = \; \Int{}{}  u_{\la}(\bs{w},  \bs{z} \mid \bs{y}_N)
\Psi_{ \bs{w} ; \ov{\bs{y}}_N-\ov{\bs{z}} }(\bs{x}_1) 
\Psi_{\bs{z} ; \veps-\ov{\bs{y}}_N+\ov{\bs{z}} }(\bs{x}_2) \;  \varpi(\bs{w},  \bs{z} \mid \bs{y}_N) 
\cdot \pl{a=1}{r-1}\dd w_a  \cdot \pl{a=r}{N-1}\dd z_a  \;,  
\enq
where we have set 
\bem
\hspace{-1cm} u_{\la}(\bs{w},  \bs{z} \mid \bs{y}_N) \; = \; \sul{p=1}{r-1} \sul{q=r}{N-1} (-1)^r
 \pl{ \substack{ a= r  \\ \not= q } }{N-1} \bigg\{ \f{ \la - z_a  }{ z_q - z_a} \bigg\}
 \pl{ \substack{a=1 \\ \not= p } }{r-1} \bigg\{ \f{ \la - w_a }{ w_p - w_a } \bigg\}
 \pl{ \substack{ b=r \\ \not= q} }{ N-1}  (z_b-w_p -i\hbar)  
  \pl{\substack{ a=1 \\ \not= p } }{r-1}  \bigg\{ \f{ 1}{  z_q - w_a } \bigg\}
\cdot \pl{b=1}{N} \bigg\{ \f{ z_q - y_b }{ y_b - w_p -i\hbar } \bigg\}   \\
\; + \; \big( \la- \ov{\bs{y}}_N + \ov{\bs{w}} + \ov{\bs{z}} + i\hbar \big) \sul{p=1}{r-1} (-1)^r 
 \pl{ \substack{ a=1  \\ \not= p } }{ r-1}(\la - w_a)   \pl{ a=r }{ N-1}(\la - z_a) 
\cdot    \f{  \pl{b=r}{N-1}(z_b - w_p -i\hbar) }{  \pl{ b=1 }{ N } \big(y_b - w_p - i \hbar \big) } 
  \cdot \pl{ \substack{a=1  \\ \not= p} }{r-1} \bigg\{ \f{1}{ w_p - w_a} \bigg\} 
 \\
\; + \; \sul{ \substack{ p, \ell   \\ p \not= \ell} }{r-1}
\f{ - \pl{  a=r  }{ N-1}(\la - z_a)  }{ (w_{\ell} - w_{p})( w_{\ell} - w_{p} + i \hbar ) }
\pl{ \substack{ a=1  \\ \not= p, \ell  } }{ r-1} \bigg\{ \f{   (\la - w_a)     }{(w_{\ell}-w_a)(w_{p}-w_a) } \bigg\}
\pl{b=r}{N-1} \bigg\{ \f{z_b - w_{\ell} -i\hbar }{ z_b - w_p  }  \bigg\} 
\pl{b=1}{N}  \bigg\{ \f{ y_b - w_p  }{y_b - w_{\ell} -i\hbar }  \bigg\}   
 \\
\; - \; \sul{p=1}{r-1}  \pl{ \substack{a=1 \\ \not= p} }{ r-1 } \bigg\{ \f{1}{ w_p - w_a} \bigg\}
\cdot \pl{ \substack{ a=1  \\ \not= p} }{r-1} (\la - w_a)
\pl{ a=r   }{N-1} (\la - z_a)   \sul{\eps= \pm }{} \pl{a=1}{r-1} \bigg\{  \f{1}{w_p - w_a +i \eps \hbar } \bigg\} \;. 
\label{ecriture de u lambda complete}
\end{multline}

Then, taking into account that 
\beq
\sul{p=1}{N} (-i)^{N+1}   \pl{ \substack{ a=1 \\ \not= p} }{N} \bigg\{ \f{\la-y_a}{y_p-y_a} \bigg\} \cdot 
\varpi(\bs{w},  \bs{z} \mid \bs{y}_N - i \hbar \bs{e}_p)  \; = \; 
		\varpi(\bs{w},  \bs{z} \mid \bs{y}_N)\; \cdot \; f_{\la}(\bs{w},  \bs{z} \mid \bs{y}_N)
\enq
with
\beq
f_{\la}(\bs{w},  \bs{z} \mid \bs{y}_N) \; = \; - \sul{p=1}{N} \;  \; 
\f{ \pl{a=r}{N-1}(y_p-z_a) }{ \pl{a=1}{r-1} (y_p - w_a -i\hbar)  }  \cdot 
 \pl{ \substack{ a=1 \\  \not= p} }{N} \bigg\{ \f{\la-y_a}{y_p-y_a} \bigg\}  
\enq
we infer that the action of the operator $D_{1, N+1}(\la)$ will take the form \eqref{ecriture action D sur Psi}
as soon as we prove that 
\beq
r_{\la}^{(2)}(\bs{w},  \bs{z} \mid \bs{y}_N)  \; = \; f_{\la}(\bs{w},  \bs{z} \mid \bs{y}_N) \; - \; u_{\la}(\bs{w},  \bs{z} \mid \bs{y}_N)
\quad \e{vanishes} \;. 
\enq

 Since it is a polynomial of degree $N-1$,
it is enough to prove that it vanishes at $N$ points. Below, we show that, indeed, it vanishes at 
\begin{itemize}
\item $\la = w_p$ with $p=1,\dots, r-1$  ; 
\item $\la = z_p$ with $p=r,\dots, N-1$  ; 
\end{itemize}
and that it behaves as $r_{\la}^{(2)}(\bs{w},  \bs{z} \mid \bs{y}_N) =\e{O}(\la^{N-1})  $ in the $ \la \tend \infty$ regime. 

\subsubsection*{ $\bullet$ Behaviour at $\infty$}
It follows from an immediate inspection that 
\beq
u_{\la}(\bs{w},  \bs{z} \mid \bs{y}_N) \; = \;   \la^{N-1} \kappa_{N-1} \; + \;  \e{O}\big( \la^{N-2} \big) 
\quad \e{with} \quad
\kappa_{N-1} \; = \; (-1)^r \sul{p=1}{r-1}  
 \Bigg\{  \f{\pl{b=r}{N-1}(z_b - w_p -i\hbar)   }{\pl{ b=1 }{ N } ( y_b - w_p - i \hbar) } \Bigg\} 
  \cdot \pl{ \substack{a=1  \\ \not= p} }{r-1} \bigg\{ \f{1}{ w_p - w_a}  \bigg\} \; . 
\enq
The leading coefficient can be recast as a contour integral which, then, can be computed by taking the residues outside of the 
integration contour. This yields
\beq
\kappa_{N-1} \; = \; (-1)^r \Oint{  \msc{C}\big( \{w_a\}_1^{r-1} \big)  }{}  
 \Bigg\{  \f{\pl{b=r}{N-1}(z_b - \om -i\hbar)   }{\pl{ b=1 }{ N } ( y_b - \om - i \hbar) } \Bigg\} 
  \cdot \pl{ a=1   }{r-1} \bigg\{ \f{1}{ \om - w_a} \bigg\} \cdot \f{  \dd \om }{2i\pi } \; = \; 
(-1)^r  \sul{\ell=1}{N}  \f{ \pl{b=r}{N-1}(z_b - y_{\ell} ) }{ \pl{ a=1   }{r-1} ( y_{\ell} - w_a - i\hbar)  }  \; \cdot  \; 
  \pl{ \substack{ b=1 \\ \not= \ell } }{N}  \bigg\{ \f{1}{ y_b - y_{\ell} }  \bigg\}  \;.   
\enq
There $ \msc{C}\big( \{w_a\}_1^{r-1} \big)$ refers to a counterclockwise loop of index $1$ around the points $w_a$ , $a=1, \dots, r-1$
but not encircling any other singularity of the integrand. 

It is immediate to see from here that, indeed, 
\beq
\kappa_{N-1} \; = \; \lim_{\la \tend \infty} \Big\{ \la^{1-N} f_{\la}(\bs{w},  \bs{z} \mid \bs{y}_N)  \Big\} \;. 
\enq

\subsubsection*{$\bullet$ Interpolation at $\la = z_q$ with $q=r,\dots,N-1$}

Setting $\la = z_q$ with $q=r,\dots,N-1$ in \eqref{ecriture de u lambda complete} leads to 
\bem
u_{z_q}(\bs{w},  \bs{z} \mid \bs{y}_N) \; = \; 
\sul{p=1}{r-1}  (-1)^r
 \pl{ \substack{a=1 \\ \not= p } }{r-1} \bigg\{ \f{1 }{ w_p - w_a } \bigg\}
  \cdot  \pl{ \substack{ b=r \\ \not= q} }{ N-1}  (z_b-w_p -i\hbar)  
\cdot \pl{b=1}{N} \bigg\{ \f{ z_q - y_b }{ y_b - w_p -i\hbar }  \bigg\} \\
\; = \;  (-1)^r \Oint{ \Ga\big( \{ w_a \}_1^{r-1} \big) }{}  
 \pl{ a=1  }{r-1} \bigg\{ \f{1 }{ \om - w_a } \bigg\}
\cdot  \pl{ \substack{ b=r \\ \not= q} }{ N-1}  (z_b-\om  -i\hbar)  
\cdot \pl{b=1}{N} \bigg\{ \f{ z_q - y_b }{ y_b - \om  -i\hbar } \bigg\}  \cdot \f{  \dd \om }{2i\pi }  \\
\; =\; (-1)^r   \sul{\ell=1}{N} 
 \pl{ a=1  }{r-1} \bigg\{ \f{1 }{ y_{\ell} - w_a - i \hbar  } \bigg\}
 \pl{ \substack{ b=r \\ \not= q} }{ N-1}  (z_b-y_{\ell} )    \cdot \pl{b=1}{N}(z_q - y_b)  \cdot 
 \pl{ \substack{ b=1 \\ \not= \ell} }{N} \bigg\{ \f{ 1 }{ y_b - y_{\ell} } \bigg\}  \\
 \; =\; - \sul{\ell=1}{N} 
  \f{\pl{ b=r }{ N-1}  (y_{\ell} -z_b )  }{ \pl{ a=1  }{r-1} \big( y_{\ell} - w_a - i \hbar \big)   } 
 \cdot  \pl{ \substack{ b=1 \\ \not= \ell} }{N} \bigg\{ \f{ z_q - y_b  }{  y_{\ell} -y_b  }  \bigg\}  
 \; = \; f_{z_q}(\bs{w},  \bs{z} \mid \bs{y}) \;. 
\end{multline}

\subsubsection*{$\bullet$ Interpolation at $\la = w_p$ with $p=1,\dots,r-1$}

A straightforward calculation shows that 
\beq
u_{w_q}(\bs{w},  \bs{z} \mid \bs{y}_N) \; = \; \mc{L}_1 + \dots +\mc{L}_5 \;,
\enq
where we agree upon
\beq
\mc{L}_1 \; = \;  (-1)^r \sul{q=r}{N-1} 
 \pl{ \substack{ a= r  \\ \not= q } }{N-1} \bigg\{ \f{ w_p- z_a  }{ z_q - z_a} \bigg\}
 \pl{ \substack{ b=r \\ \not= q} }{ N-1}  (z_b-w_p -i\hbar)  
  \pl{\substack{ a=1 \\ \not= p } }{r-1}  \bigg\{ \f{ 1}{  z_q - w_a } \bigg\}
\cdot \pl{b=1}{N}\bigg\{ \f{ z_q - y_b }{ y_b - w_p -i\hbar } \bigg\} \;,
\enq
\beq
\mc{L}_2 \; = \; (-1)^r    
\f{\pl{ a=r }{ N-1} \Big\{ (w_p - z_a) (z_a - w_p -i\hbar)    \Big\}   }
		{    \pl{ b=1 }{ N } (y_b - w_p - i \hbar)  } 
\cdot  \big( w_p- \ov{\bs{y}}_N + \ov{\bs{w}} + \ov{\bs{z}} + i\hbar \big)   \;, 
\enq
\beq
\mc{L}_3 \; = \;\pl{  a=r  }{ N-1}(w_p - z_a) \sul{ \substack{  \ell =1  \\ \ell \not= p} }{r-1}
\f{ - 1 }{ (w_{\ell} - w_{p})( w_{\ell} - w_{p} + i \hbar ) }
\pl{ \substack{ a=1  \\ \not= p, \ell  } }{ r-1} \bigg\{ \f{1}{ w_{\ell}-w_a } \bigg\}
\pl{b=r}{N-1} \bigg\{ \f{z_b - w_{\ell} -i\hbar }{ z_b - w_p  }  \bigg\} 
\pl{b=1}{N}  \bigg\{ \f{ y_b - w_p  }{y_b - w_{\ell} -i\hbar }  \bigg\} \; , 
\enq
\beq
\mc{L}_4 \; = \;\pl{  a=r  }{ N-1}(w_p - z_a) \sul{ \substack{  \ell =1  \\ \ell \not= p} }{r-1}
\f{ - 1 }{ (w_{p} - w_{\ell})( w_{p} - w_{\ell} + i \hbar ) }
\pl{ \substack{ a=1  \\ \not= p, \ell  } }{ r-1} \bigg\{ \f{1}{w_{\ell}-w_a } \bigg\}
\pl{b=r}{N-1} \bigg\{ \f{z_b - w_{p} -i\hbar }{ z_b - w_{\ell}  }  \bigg\} 
\pl{b=1}{N}  \bigg\{ \f{ y_b - w_{\ell}  }{y_b - w_{p} -i\hbar }  \bigg\} \; , 
\enq
and
\beq
\mc{L}_5 \; = \; \; - \; \pl{ a=r   }{N-1} (w_p - z_a)   \;  \cdot \; 
\sul{\eps= \pm }{} \pl{a=1}{r-1} \bigg\{  \f{1}{w_p - w_a +i \eps \hbar } \bigg\} \; . 
\enq

\noindent Most of the sums can be re-expressed in terms of contour integrals. Namely, one has 
\beq
\mc{L}_1 \; = \;  (-1)^r \f{  \pl{a=r}{N-1} \Big\{  (w_p- z_a) (z_a-w_p-i\hbar) \Big\}  }{ \pl{b=1}{N} (y_b - w_p -i\hbar )  }  
 \cdot \mc{S}_1
\enq
where
\beq 
\mc{S}_1 \; = \;   \sul{q=r}{N-1} \f{-1}{ z_q - w_p -i\hbar }  \cdot 
 \pl{ \substack{ a= r  \\ \not= q } }{N-1} \bigg\{ \f{ 1  }{ z_q - z_a} \bigg\} \cdot 
 \pl{a=1  }{r-1}  \bigg\{ \f{ 1}{  z_q - w_a } \bigg\} \cdot \pl{b=1}{N} ( z_q - y_b) \;. 
\enq
Thus it follows that 
\bem
\mc{S}_1 \; = \; - \Oint{ \msc{C}\big( \{ z_a \}_{r}^{N-1} \big) }{} \hspace{-3mm}
\f{1}{  \om - w_p -i\hbar }  \cdot  \pl{  a= r  }{N-1} \bigg\{ \f{ 1  }{ \om - z_a} \bigg\}
\cdot    \pl{ a=1 }{r-1}  \bigg\{ \f{ 1}{  \om - w_a } \bigg\}
\cdot \pl{b=1}{N} ( \om- y_b) \cdot \f{\dd \om }{2i\pi}  \\
\; = \;    \f{  \pl{b=1}{N} ( w_p - y_b + i \hbar) }
{ \pl{  a= r  }{N-1} (w_p - z_a + i\hbar)   \pl{ a=1 }{r-1}  ( w_p - w_a +i\hbar  ) }
\; + \; \sul{\ell=1}{r-1} \f{  \pl{b=1}{N} ( w_{\ell} - y_b)  }{  w_{\ell} - w_p -i\hbar }  
\cdot  \pl{  a= r  }{N-1} \bigg\{ \f{ 1  }{ w_{\ell} - z_a} \bigg\}
\cdot    \pl{ \substack{ a=1 \\ \not= \ell } }{r-1}  \bigg\{ \f{ 1}{  w_{\ell} - w_a } \bigg\} \\
\; - \; \big( w_p + i\hbar + \ov{\bs{w}} + \ov{\bs{z}} - \ov{\bs{y}}_N \big) \;, 
\end{multline}
where we have taken the integral by computing the residues at the poles lying outside of the contour. 
Note that there was a non-vanishing residue at $\infty$. 

As a consequence, we get 
\bem
\mc{L}_1 + \mc{L}_2 + \mc{L}_5 \; = \; 
 \f{  \pl{a=r}{N-1} \Big\{  (w_p- z_a) (z_a-w_p-i\hbar) \Big\}  }{ \pl{b=1}{N} (y_b - w_p -i\hbar )  } 
 \sul{\ell=1}{r-1} \f{  \pl{b=1}{N} ( y_b-  w_{\ell})  }{  w_{\ell} - w_p -i\hbar }  
\cdot  \pl{  a= r  }{N-1} \bigg\{ \f{ 1  }{  z_a - w_{\ell} } \bigg\}  \cdot 
\pl{ \substack{ a=1 \\ \not= \ell}  }{ r }  \bigg\{ \f{ 1  }{  w_{\ell} - w_a } \bigg\} \\ 
\; - \; \f{ \pl{a=r}{N-1} (w_p - z_a)  }{ \pl{a=1}{r-1} (w_p - w_a -i\hbar)  } \;. 
\end{multline}

We now rewrite $\mc{L}_3$:
\bem
\mc{L}_3 \; = \;  -\f{i}{\hbar}  
\pl{ \substack{ a=1  \\ \not=  p  } }{ r-1} \bigg\{ \f{1}{w_{p}-w_a } \bigg\}
\cdot  \pl{b=r}{N-1} \Big\{ w_{p}- z_b +i\hbar  \Big\} 
\cdot  \pl{b=1}{N}  \bigg\{ \f{ y_b - w_p  }{y_b - w_{\ell} -i\hbar }  \bigg\}  \\
\; + \;  \Oint{ \msc{C}\big( \{w_a\}_1^N \big) }{}  
\f{ \pl{b=r}{N-1} (\om- z_b +  i\hbar  )  }{  w_{p} - \om - i \hbar  }
\cdot  \pl{ a=1  }{ r-1} \bigg\{ \f{1}{\om - w_a } \bigg\}
\cdot  \pl{b=1}{N}  \bigg\{ \f{ y_b - w_p   }{y_b - \om  -i\hbar }  \bigg\} \cdot \f{ \dd \om }{ 2i\pi}  \\
 \; = \;  -\f{i}{\hbar}  
\pl{ \substack{ a=1  \\ \not=  p  } }{ r-1} \bigg\{ \f{1}{w_{p}-w_a } \bigg\}
 \cdot  \pl{b=r}{N-1} \Big\{ w_{p}- z_b +i\hbar  \Big\} 
 \cdot  \pl{b=1}{N}  \bigg\{ \f{ y_b - w_p  }{y_b - w_{\ell} -i\hbar }  \bigg\}  
\; + \; \f{  \pl{b=r}{N-1} (w_p- z_b)  }{  \pl{ a=1  }{ r-1} (w_p - w_a -i\hbar)  } \\
\; + \;  \sul{ \ell= 1 }{ N } \f{ \pl{b=r}{N-1} (y_{\ell}- z_b   )  }{  w_{p} - y_{\ell}  }
\cdot   \f{\pl{b=1}{N}  (y_b - w_p )   }{\pl{ a=1  }{ r-1} (y_{\ell} - w_a  - i \hbar) } 
\cdot  \pl{ \substack{  b=1 \\ \not= \ell}  }{ N } \bigg\{ \f{ 1  }{y_b - y_{\ell} }  \bigg\}   \; . 
\end{multline}

Finally, it is readily seen that 
\bem
\mc{L}_4 \; = \; \f{i}{\hbar} \pl{b=r}{N-1} (w_p  - z_{b} + i\hbar)
\cdot  \pl{ \substack{ a=1  \\ \not= p  } }{ r-1} \bigg\{ \f{1}{ w_{p}-w_a } \bigg\}
\cdot  \pl{b=1}{N}  \bigg\{ \f{ y_b - w_{p}  }{y_b - w_{p} -i\hbar }  \bigg\}  \\
\; +\; \pl{  a=r  }{ N-1}(w_p - z_a) \sul{ \ell =1    }{r-1}
\f{  1 }{  w_{p} - w_{\ell} + i \hbar  }
\pl{ \substack{ a=1  \\ \not= \ell  } }{ r-1} \bigg\{ \f{1}{w_{\ell}-w_a } \bigg\}
\cdot  \pl{b=r}{N-1} \bigg\{ \f{z_b - w_{p} -i\hbar }{ z_b - w_{\ell}  }  \bigg\} 
\cdot  \pl{b=1}{N}  \bigg\{ \f{ y_b - w_{\ell}  }{y_b - w_{p} -i\hbar }  \bigg\} \;. 
\end{multline}
Hence, adding up all the partial representations for the $\mc{L}_k$'s, we get that
\beq
\sul{p=1}{5} \mc{L}_{p} \; = \; - \sul{ \ell= 1 }{ N }
\f{ \pl{b=r}{N-1} (y_{\ell}- z_b   )  }{ \pl{ a=1  }{ r-1} (y_{\ell} - w_a  - i \hbar)   }
\cdot  \pl{ \substack{  b=1 \\ \not= \ell}  }{ N } \bigg\{ \f{ y_b - w_p    }{y_b - y_{\ell} }  \bigg\}   
\; = \;  f_{w_p } (\bs{w}, \bs{z} \mid \bs{y} ) \; . 
\enq

The form of the action of the operator $A_{1,N+1}(\la)$ can be readily inferred by evaluating 
at $ \la = y_k $
the action of the below form of the quantum determinant relation 
\beq
A_{1,N+1}(\la-i\hbar) D_{1,N+1}(\la) \; - \;  C_{1,N+1}(\la-i\hbar) B_{1,N+1}(\la) \; = \; 1  \;. 
\label{ecriture determinant quantique dans preuve rep int fct Whittaker}
\enq
 on $\Psi_{\bs{y}_N,\veps}(\bs{x}_{N+1})$. This removes the contribution coming from $B_{1,N+1}(y_k)$
whereas the action of $D_{1,N+1}(y_k)$ is known. This allows one to interpolate the action of $A_{1,N+1}(\la)$
at $N$ points. Furthermore, one has the following asymptotic behaviour 
\beq
A_{1,N+1}(\la) \; = \; \la^{N+1} - \la^{N}  \Big( \sul{a=1}{N+1}p_a \Big)  \; + \; \e{O} \big( \la^{N-2} \big)
\enq
Since, it is readily seen from the induction hypothesis that 
\beq
 \Big( \sul{a=1}{N+1}p_a \Big)  \cdot \Psi_{\bs{y}_N,\veps}(\bs{x}_{N+1}) \; = \; 
 \veps  \cdot \Psi_{\bs{y}_N,\veps}(\bs{x}_{N+1}) \;, 
\enq
the full form for the action of the operator $A_{1,N+1}(\la)$ follows.

\subsection*{$\bullet$ Independence on the splitting parameter $r$}

It follows from the joint results of \cite{GerasimovKharchevLebedevRepThandQISM,GoodmannWallachQuantumTodaIII,KostantIdentificationOfEigenfunctionsOpenTodaAndWhittakerVectorsMoreDeep}
or, more directly, from \cite{KozUnitarityofSoVTransform}  that, up to a multiplicative constant, 
there exist a unique function $f_{\bs{y}_N;\veps}(\bs{x}_{N+1})$
such that 
\beq
B_{1,N+1}(\la) \cdot f_{\bs{y}_N;\veps}(\bs{x}_{N+1}) \; = \;  \pl{a=1}{N}(\la-y_a) \cdot  f_{\bs{y}_N;\veps + i\hbar }(\bs{x}_{N+1})
\qquad \e{for} \; \; \e{any} \qquad \la \in \R \;. 
\label{ecriture equation pour fct propre B}
\enq
Let $\Psi^{(r)}_{\bs{y}_N;\veps}(\bs{x}_{N+1})$ be as constructed through \eqref{ecriture rep int fct propre B}
with a splitting $r$. Since, independently of the value of $r$ these functions satisfy \eqref{ecriture equation pour fct propre B},
there exists a constant $c_{r, \ell}$ such that 
\beq
\Psi^{(r)}_{\bs{y}_N;\veps}(\bs{x}_{N+1}) \; = \; c_{r, \ell} \cdot \Psi^{(\ell)}_{\bs{y}_N;\veps}(\bs{x}_{N+1}) \;. 
\enq
It remains to fix this constant by taking the $\bs{x}_{N+1} \tend \infty$ asymptotics of both expressions. 
In fact, it is enough to compute the leading asymptotic expansion in one direction of $\R^{N+1}$. 
For technical reasons, we shall focus on the following limit
\beq
x_a \tend +\infty \quad a=1,\dots, N+1 \qquad \e{such} \;  \e{that} \qquad x_{a+1}-x_{a} \tend + \infty  \; . 
\label{ecriture simplex RN+1 ou lon approche linfini}
\enq

Taken into account the independence on the splitting part in the construction of $\Psi$ functions 
in less than N variables, it follows from the results obtained in \cite{KozUnitarityofSoVTransform}, equations (A.7)-(A.8), that 
$\Psi^{(1)}_{\bs{y}_N;\veps}(\bs{x}_{N+1})$ admits the integral representation 
\beq
\Psi^{(1)}_{\bs{y}_N;\veps}(\bs{x}_{N+1}) \; = \; \ex{\f{i}{\hbar} (\veps-\ov{\bs{y}}_N) x_{N+1} }  \cdot 
\sul{ \tau \in \mf{S}_N }{} J_N\big( \bs{y}_{N;\tau} , \bs{x}_N \big) \qquad \e{with} \qquad 
\bs{y}_{N;\tau} \; = \; \big( y_{\tau(1)}, \dots, y_{\tau(N)} \big)
\label{ecriture expression Psi via J}
\enq
where
\beq
J_N\big( \bs{w}_{N}^{(0)} ; \bs{x}_N\big) \; = \; \ex{ \f{i}{\hbar} \ov{\bs{w}}_{N}^{(0)}\cdot  x_{N} }   
\pl{s=1}{N-1} \Int{ (\R-i\a_s)^{N-s} }{  } \hspace{-3mm} 
 \dd^{N-s} w^{(s)} 
\pl{s=1}{N-1}  \bigg\{ \ex{ \f{i}{\hbar}  (x_{N-s} \, - \, x_{N-s+1}) \cdot \ov{\bs{w}}_{N-s}^{(s)}  } \bigg\}
\cdot \;   \f{  \mc{W}_N \big( \big\{ \bs{w}_{N-s}^{(s)} \big\}_0^{N-1}  \big)   }
				{   \pl{s=1}{N-1} \pl{a=1}{N-s} \big( w_a^{(s)}-w_a^{(s-1)} \big)   }  \; \;,
\label{definition fonction J}
\enq
$0\, < \, \a_1 \, < \, \dots \, <  \, \a_{N-1} $ and 
\beq
\mc{W}_N \Big( \big\{ \bs{w}_{N-s}^{(s)} \big\}_0^{N-1}  \Big) \; = \;
  \f{   \hbar^{-\f{\hbar}{i}(N-1)\ov{\bs{w}}_N^{(0)} }  }{ \pl{b>a}{N} \big( w_a^{(0)} - w_b^{(0)} \big)  }  
\cdot \pl{s=1}{N-1} \Bigg\{ \f{ (-i\hbar)^{(N-s)} \hbar^{ 2\f{\hbar}{i} \ov{\bs{w}}^{(p)}_{N-s}  }  }{ (2 i \pi )^{N-s} }  \Bigg\}
 \cdot 
\pl{s=1}{N-1} \Bigg\{ \f{ \pl{a=1}{N-s} \pl{b=1}{N-s+1} \Ga\Big( \f{ w_b^{(s-1)} - w_a^{(s)} }{i\hbar} + 1 \Big)  }
{ \pl{a \not= b}{N-s} \Ga\Big( \f{ w_a^{(s)} - w_b^{(s)} }{i\hbar} + 1 \Big)  } \Bigg\}
  \;. 
\enq

The leading $\bs{x}_{N} \tend \infty$ asymptotic behaviour of $J_N\big( \bs{w}_{N}^{(0)} ; \bs{x}_N\big)$
can be extracted from \eqref{definition fonction J} by appropriate shifts (to the upper or lower half-planes) 
of the integration contours as described in \cite{KozUnitarityofSoVTransform}. 
For the limit of interest \eqref{ecriture simplex RN+1 ou lon approche linfini}, it is enough to move all contours
to the upper half-plane. The sole contribution that will not lead to exponentially small corrections in respect to  
$x_{a+1}-x_a$, for some $a=1,\dots, N-1$, corresponds to computing the residues at \textit{all} contiguous poles \textit{ie}. at 
\beq
\bs{w}_{N-1}^{(1)} \; = \; \bs{w}^{(0)}_{N-1} \qquad \e{and} \; \e{further} \; \e{successively} \qquad
\bs{w}^{(s)}_{N-s} \; = \; \bs{w}^{(0)}_{N-s} \;. 
\enq
It is readily seen that this leads, all-in-all, to the following $\bs{x}_{N+1} \tend \infty$
asymptotics
\beq
J_N\big( \bs{w}_{N}^{(0)} ; \bs{x}_N\big) \; = \;  \pl{s=1}{N}\hbar^{\f{i}{\hbar} (N+1-2s)w_s^{(0)} }
\cdot \pl{a=1}{N} \ex{ \f{i}{\hbar} x_s w_s^{(0)}  } \cdot \pl{a<b}{N}\Ga\Big( \f{ w_b^{(0)}-w_a^{(0)} }{i\hbar} \Big)
 \; + \; \e{O}\big( \sul{s=1}{N-1} [x_{s+1}-x_s]^{-\infty} \big)
\enq
in the direction \eqref{ecriture simplex RN+1 ou lon approche linfini}. Thus, the asymptotics of 
$\Psi^{(1)}_{\bs{y}_N;\veps}(\bs{x}_{N+1})$ in the very same direction read 
\beq
\Psi^{(1)}_{\bs{y}_N;\veps}(\bs{x}_{N+1}) \; = \;  \ex{\f{i}{\hbar}(\veps - \ov{\bs{y}}_N)x_{N+1} }
\sul{ \tau \in \mf{S}_N }{} \pl{s=1}{N}\hbar^{\f{i}{\hbar} (N+1-2s)y_{\tau(s)} }
\cdot \pl{a=1}{N} \ex{ \f{i}{\hbar} x_s y_{\tau(s)}  } \cdot \pl{a<b}{N}\Ga\Big( \f{ y_{\tau(b)}- y_{\tau(a)} }{i\hbar} \Big)
 \; + \; \e{O}\big( \sul{s=1}{N-1} [x_{s+1}-x_s]^{-\infty} \big) \;. 
\enq

By the induction hypothesis, for any value of $r$, 
the integrand in \eqref{ecriture rep int fct propre B} can be recast by means of the lower number of 
variables version of formula \eqref{ecriture expression Psi via J}, namely 
\bem
\Psi^{(r)}_{\bs{y}_N;\veps}(\bs{x}_{N+1}) \; = \;  \ex{\f{i}{\hbar}(\veps - \ov{\bs{y}}_N)x_{N+1} } \cdot 
\ex{\f{i}{\hbar} \ov{\bs{y}}_N x_r} \cdot (r-1)!(N-r)! \\
\times \Int{ \msc{C}_{r-1;N-r} }{} \hspace{-2mm} 
J_{r-1}\big(\bs{w};\bs{x}_{r-1} \big) \cdot J_{N-r}\big( \bs{z}; \wt{\bs{x}}_{N-r}  \big) \cdot 
\ex{ -\f{\hbar}{i}(\ov{\bs{w}}+\ov{\bs{z}})x_r }
\cdot \varpi\big(\bs{w}, \bs{z} \mid \bs{y}_N  \big) \cdot \pl{a=1}{r-1} \dd w_a \cdot \pl{a=r}{N-1} \dd z_a \;, 
\label{ecriture rep int Psi r}
\end{multline}
where we agree upon 
\beq
\bs{x}_{r-1} \; = \; \big( x_1, \dots, x_{r-1} \big) \qquad \e{and} \qquad 
\wt{\bs{x}}_{N-r} \; = \; \big( x_r, \dots, x_{N} \big) \;. 
\enq

In order to extract the leading $\e{O}(1)$, $\bs{x}_{N+1}\tend \infty$ as in \eqref{ecriture simplex RN+1 ou lon approche linfini}, contributions of
\eqref{ecriture rep int Psi r} one should move the $\bs{w}$-integration to the upper half-plane and the 
$\bs{z}$-integration slightly to the lower half-plane. The sole contribution not leading to some 
exponentially small term corresponds to taking the residues at 
\beq
w_a = y_{\sg(a)}  \quad \e{for} \quad a=1,\dots,r-1 \qquad \e{and} \qquad
z_a \; = \; y_{\sg(a+1)} \quad \e{for} \quad a=r,\dots,N-1  
\enq
when $\sg$ runs through $\mf{S}_N$. A straightforward computation then shows that for 
$\bs{x}_{N+1} \tend \infty $ as in  \eqref{ecriture simplex RN+1 ou lon approche linfini}, 
\beq
\Psi^{(r)}_{\bs{y}_N;\veps}(\bs{x}_{N+1}) \; = \;  \ex{\f{i}{\hbar}(\veps - \ov{\bs{y}}_N)x_{N+1} }
\sul{ \tau \in \mf{S}_N }{} \pl{s=1}{N}\hbar^{\f{i}{\hbar} (N+1-2s)y_{\tau(s)} }
\cdot \pl{a=1}{N} \ex{ \f{i}{\hbar} x_s y_{\tau(s)}  } \cdot \pl{a<b}{N}\Ga\Big( \f{ y_{\tau(b)}- y_{\tau(a)} }{i\hbar} \Big)
 \; + \; \e{O}\big( \sul{s=1}{N-1} [x_{s+1}-x_s]^{-\infty} \big) \;. 
\enq
Hence $c_{r,\ell}=1$ and the independence of $\Psi_{\bs{y}_N, \veps}(\bs{x}_{N+1})$ on the splitting $r$ used in the 
integral representation \eqref{ecriture rep int fct propre B} follows.  \qed




\section{Proof of Proposition \ref{proposition action operateur Or sur fct propres B}}
\label{Appendix Proof prop action op Or}

The starting point for computing the action of the operator $\mc{O}_r(\la)$ on $\Psi_{\bs{y}_N,\veps}(\bs{x}_{N+1})$ is to use 
the integral representation \eqref{ecriture rep int fct propre B} for $\Psi_{\bs{y}_N,\veps}(\bs{x}_{N+1})$. 
This has the advantage of allowing one to act directly, under the integral sign,
\begin{itemize}
\item with the product of position operators on $\Psi_{\bs{w} , \ov{\bs{y}}_N - \ov{\bs{z}}  }(\bs{x}_1)  $
by means of  \eqref{ecriture proprietes action ex x sur fct propre B}; 
\item   with the operator $D_{r+1, N+1}(\la)$ on $\Psi_{\bs{z} , \veps - \ov{\bs{y}}_N + \ov{\bs{z}} }(\bs{x}_2) $
by means of \eqref{ecriture action D sur Psi}. 
\end{itemize}
Shifting the integration contours along the lines described in Appendix \ref{Appendix Proof Prop action ABD}
recasts the action in the form 
\bem
\mc{O}_r(\la)  \cdot \Psi_{\bs{y}_N,\veps}(\bs{x}_{N+1})
\; = \; \Int{ \msc{C}_{r-1,N-r}}{}  \Psi_{\bs{w} , \ov{\bs{y}}_N - \ov{\bs{z}} - i (r+1) \hbar }(\bs{x}_1)
\Psi_{\bs{z} , \veps - \ov{\bs{y}}_N + \ov{\bs{z}} + i (r+1) \hbar }(\bs{x}_2)  \\
\times \; h_{\la}(\bs{w} ,  \bs{z}  \mid \bs{y}_N)
\cdot \varpi(\bs{w} ,  \bs{z}  \mid \bs{y}_N)   \pl{a=1}{r-1} \dd w_a \cdot \pl{a=r}{N-1} \dd z_a \;, 
\end{multline}
where we have set 
\beq
h_{\la}(\bs{w} ,  \bs{z}  \mid \bs{y}_N) \; = \; 
 (i)^{r(r-1)}  \pl{a=1}{r-1} \bigg\{ \pl{b=r}{N-1} (z_b-w_a-i\hbar) \cdot \pl{b=1}{N} \f{1}{ y_b-w_a-i\hbar } \bigg\} \times 
\sul{p=r}{N-1}  \f{ \pl{b=1}{N}(z_p-y_b)  }{  \pl{a=1}{r-1} (z_p-w_a-i\hbar) }  
\pl{ \substack{ a=r \\ \not= p} }{ N-1 } \bigg( \f{\la-z_a}{z_p-z_a} \bigg) \; . 
\label{definition fonction u lambda}
\enq

Similarly, one can re-express the form of the action given in  \eqref{ecriture action operateur O r sur Psi} by moving the multiple sum
under the integral sign leading to 
\bem
- \sul{ \substack{  \mc{I}_N =  \sg \cup  \ov{\sg} \\ \#\sg = r+1} }{}  \pl{ \substack{ a\in \sg \\ b\not\in \sg }}{} \bigg\{ \f{- i}{y_a-y_b} \bigg\}
\cdot  \pl{ b\not\in \sg }{} (\la-y_b)  \cdot \Psi_{\bs{y}_N-i\hbar \sul{a \in \sg }{} \bs{e}_a,\veps}(\bs{x}_{N+1}) \\
= \Int{ \msc{C}_{r-1,N-r}}{}  \Psi_{\bs{w} , \ov{\bs{y}}_N - \ov{\bs{z}} - i (r+1) \hbar }(\bs{x}_1)
\Psi_{\bs{z} , \veps - \ov{\bs{y}}_N + \ov{\bs{z}} + i (r+1) \hbar }(\bs{x}_2)  
\cdot  t_{\la}(\bs{w} ,  \bs{z}  \mid \bs{y}_N)
\cdot \varpi(\bs{w} ,  \bs{z}  \mid \bs{y}_N)   \pl{a=1}{r-1} \dd w_a \cdot \pl{a=r}{N-1} \dd z_a \;, 
\end{multline}
 where 
\beq
t_{\la}(\bs{w} ,  \bs{z}  \mid \bs{y}_N) \; = \;  - (i)^{(r+1)(r-2)} 
\sul{ \substack{ \mc{I}_N =  \sg \cup  \ov{\sg} \\ \#\sg = r+1} }{}  
\pl{ \substack{ a\in \sg \\ b\not\in \sg }}{} \bigg\{ \f{- 1 }{y_a-y_b} \bigg\}
\cdot  \pl{ b\not\in \sg }{} (\la-y_a)  \cdot \pl{b\in \sg}{} \bigg\{ \pl{a=r}{N-1} (z_a - y_b)  
\cdot \pl{a=1}{r-1} \f{ 1 }{ y_b-w_a-i\hbar }  \bigg\}  \;. 
\label{definition finale fonction f lambda}
\enq

Note that, the multiple sum in the definition of $t_{\la}(\bs{w},\bs{z} \mid \bs{y}_N)$ 
can be re-cast in terms of a single $(r+1)$-fold contour integral 
\bem
f_{\la}(\bs{w} ,  \bs{z}  \mid \bs{y}_N) \; = \; - \f{ (i)^{(r+1)r} }{ (r+1)! } \Oint{ \msc{C}( \{ y_a \}_1^N ) }{}  
\f{  \pl{b=1}{N} (\la - y_a) }{  \pl{a=1}{r+1} (\la- s_a) } \cdot \pl{a\not=b}{r+1} (s_a - s_b)  \pl{a=1}{r+1} \pl{b=1}{N} \Big[ \f{1}{y_b-s_a}\Big] \\
\times 
\pl{b=1}{r+1}  \bigg\{  \pl{a=r}{N-1} (z_a-s_b) \pl{a=1}{r-1} \Big[ \f{1}{ s_b - w_a -i\hbar} \Big] \bigg\} \cdot \f{ \dd^{r+1}s }{ (2i\pi)^{r+1} } \;. 
\label{ecriture representation integrale pour f lambda}
\end{multline}

The contour $\msc{C}(\{y_a\}_1^N )$ appearing above is a counterclockwise loop of index 1 around each of the $y$'s that does 
not surround any of the other poles in the integration variables $s_a$, with $a=1,\dots, r+1$.

 As a consequence, one gets that equations \eqref{ecriture action operateur O r sur Psi} 
will follow as soon as we have shown that 
\beq
r^{(3)}_{\la}(\bs{w} ,  \bs{z}  \mid \bs{y}_N) \; = \; h_{\la}(\bs{w} ,  \bs{z}  \mid \bs{y}_N)
 \; - \;  t_{\la}(\bs{w} ,  \bs{z}  \mid \bs{y}_N)  
\enq
vanishes. 

Notice that the functions defined in \eqref{definition fonction u lambda} and  \eqref{definition finale fonction f lambda} are both 
polynomials in $\la$ of degree $N-1-r$. Hence, it is enough to show that $r^{(3)}_{\la}(\bs{w} ,  \bs{z}  \mid \bs{y}_N)$
\begin{itemize}
\item  vanishes at the points $\la= z_p$ with $p=r,\dots, N-1$ ;
\item has $\e{O}(\la^{N-2-r})$ leading asymptotics around the point $\la = \infty$, \textit{ie}
\; $r^{(3)}_{\la}(\bs{w} ,  \bs{z}  \mid \bs{y}_N)\; = \; \e{O}(\la^{N-2-r})$. 
\end{itemize}

For this purpose, we evaluate the integral representation \eqref{ecriture representation integrale pour f lambda} 
by the residues lying outside of the original contour 
$\msc{C}( \{ y_a \}_1^N )$. This demands a little care as, individually in each variable $s_a$, the integrand behaves as 
a constant when $s_a \tend \infty$. Hence one has to take into account the contribution of the residue at $s_a= \infty$. 
Still, the explicit computation of this residue's contribution can be avoided first 
by regularizing the integral (so that it has a faster decay at infinity) and then, once all the calculations are finished, 
removing the regularization parameter. More precisely, in order to show the equality at the point  $\la= z_p$ with $p=r,\dots, N-1$ 
we regularize the integral representation for $t_{z_p}(\bs{w} ,  \bs{z}  \mid \bs{y}_N)$ in the form 
\bem
\wt{t}_p(\bs{w} ,  \bs{z}  \mid \bs{y}_N) \; = \; - \f{ (i)^{(r+1)r} }{ (r+1)! }  (w_r w_{r+1})^{r+1}  \pl{b=1}{N} (z_p - y_b)  \\
\times  \Oint{ \msc{C}( \{ y_a \}_1^N ) }{}   \cdot \pl{a\not=b}{r+1} (s_a - s_b)  \pl{a=1}{r+1} \pl{b=1}{N} \Big[ \f{1}{y_b-s_a}\Big] 
\pl{b=1}{r+1}  \bigg\{  \pl{ \substack{ a=r \\ \not= p } }{N-1} (z_a-s_b) \pl{a=1}{r+1} \Big[ \f{1}{ s_b - w_a -i\hbar} \Big] \bigg\} \cdot \f{ \dd^{r+1}s }{ (2i\pi)^{r+1} } \;. 
\label{ecriture rep int interp f at zp regularisee}
\end{multline}
The above function is built in such a way that
\beq
\wt{t}_p(\bs{w} ,  \bs{z}  \mid \bs{y}_N) \;  \limit{ w_r, w_{r+1} }{ \infty }  \;t_{z_p} (\bs{w} ,  \bs{z}  \mid \bs{y}_N) \; . 
\label{ecriture relation f tile p et f at zp}
\enq
The integral representation \eqref{ecriture rep int interp f at zp regularisee} can be easily evalued by taking the 
residues outside of the contour $\msc{C}( \{y_a\}_1^N ) $. The integrand has poles in respect to the variables $s_a$ at points 
\beq
s_a  \; = \;  w_b + i\hbar \qquad \e{with}  \qquad b=1,\dots, r+1  \;. 
\enq
Yet, because of the presence of the squared Van-der-Monde determinant, solely the residues computed at distinct points for 
distinct variables lead to non-zero contributions. 
Further, the symmetry of the integrand implies that the integral will be given by $(r+1)!$ times 
the residues of the integrand at the points
\beq
s_a \; =  \;  w_a + i\hbar \qquad \e{with}  \qquad a=1,\dots, r+1 \;.
\enq
As a consequence, 
\beq
\wt{t}_p(\bs{w} ,  \bs{z}  \mid \bs{y}_N) \; = \; - (-1)^{r+1} (i)^{r(r+1)}  \pl{b=1}{N} (z_p-y_b) 
\pl{b=1}{r+1} \pl{ \substack{ a=r \\ \not= p} }{N-1} (z_a-w_b-i\hbar)
\; \cdot \;  ( w_r w_{r+1} )^{r+1} 
\pl{a=1}{r+1} \pl{b=1}{N} \bigg\{ \f{ -1 }{ w_a - y_b +i\hbar }  \bigg\} \;. 
\enq
Thus, by taking in the $w_r, w_{r+1} \tend \infty$ limit, in virtue of \eqref{ecriture relation f tile p et f at zp}, 
\bem
f_{z_p} (\bs{w} ,  \bs{z}  \mid \bs{y}_N)  \; = \; (i)^{r(r-1)} 
\pl{b=1}{N} (z_p-y_b) \pl{b=1}{r-1} \pl{ a=r  }{N-1} (z_a-w_b-i\hbar)  \\
\times  \pl{a=1}{r-1} \pl{b=1}{N} \bigg\{ \f{ -1 }{ w_a - y_b +i\hbar } \bigg\} 
 \cdot \pl{a=1}{r-1} \bigg\{ \f{1}{z_p-w_a-i\hbar}  \bigg\} \; = \;  u_{z_p} (\bs{w} ,  \bs{z}  \mid \bs{y}_N)  \;. 
\end{multline}

As a consequence, it solely remains to show the equality of the leading asymptotics at $\la=\infty$. 
Sending $\la\tend \infty$ in \eqref{ecriture representation integrale pour f lambda} and restricting to the leading asymptotics
makes the behaviour of the 
integrand at $s_a=\infty$ even worse that in the previous case (the latter grows in this situation  linearly in $s_a$ at $s_a = \infty$). 
In thus appears convenient, for the purpose of intermediate calcualtions,
to regularize the integrand by adding three auxiliary parameters $w_r, w_{r+1} ,w_{r+2}$. The result of 
interest will then be recovered by sending the three variables to $\infty$. More precisely, we set 
\bem
\wt{t}_{\infty} (\bs{w} ,  \bs{z}  \mid \bs{y}_N) \; = \; - \f{ (i)^{(r+1)r} }{ (r+1)! }  \Big( \pl{a=r}{r+2}(-w_a) \Big)^{r+1} 
\Oint{ \msc{C}( \{ y_a \}_1^N ) }{}   \pl{a\not=b}{r+1} (s_a - s_b)  \pl{a=1}{r+1} \pl{b=1}{N} \Big[ \f{1}{y_b-s_a}\Big] \\
\times 
\pl{b=1}{r+1}  \bigg\{  \pl{  a=r  }{N-1} (z_a-s_b) \pl{a=1}{r+2} \Big[ \f{1}{ s_b - w_a -i\hbar} \Big] \bigg\} \cdot \f{ \dd^{r+1}s }{ (2i\pi)^{r+1} } \;, 
\label{ecriture rep int f reg a infty}
\end{multline}
%
%
so that 
\beq
\wt{t}_{\infty}(\bs{w} ,  \bs{z}  \mid \bs{y}_N) \;  \limit{ w_r, w_{r+1}, w_{r+2} }{ \infty }  \; 
\lim_{\la \tend \infty}  \Big\{ \la^{r+1-N } t_{\la} (\bs{w} ,  \bs{z}  \mid \bs{y}_N)  \Big\} \; . 
\label{ecriture relation f tile infty et f at zp}
\enq

The integral in \eqref{ecriture rep int f reg a infty} can be estimated by the residues outside of $\msc{C}( \{ y_a \}_1^N )$.
These are located at 
\beq
 s_a \; = \; w_b +i\hbar \qquad \e{with} \qquad b=1,\dots,r+2 \; .  
\enq
Again, due to the presence of the Van-der-Monde determiants, the only choice of residues giving non-zero contribution 
corresponds to computing the residues at 
\beq
 \{ s_a \}_1^{r+1} \;  = \; \{ w_a  + i\hbar \}_{1, \not= p} ^{r+2}  \qquad \e{for} \qquad p=1,\dots, r+2 \;. 
\enq
Each of these contributions ought to be weighted by the factor $(r+1)!$ originating from the symmetry of the integrand. 
Hence, 
\bem
\wt{t}_{\infty}(\bs{w} ,  \bs{z}  \mid \bs{y}_N) \; = \; (i)^{r(r-1)}  \Big( \pl{a=r}{r+2}(-w_a) \Big)^{r+1}  \;\; 
\sul{p=1}{r+2} \; \pl{b=1}{N} \pl{ \substack{a=1 \\ \not= p} }{r+2} \bigg\{ \f{1}{ y_b-w_a-i\hbar} \bigg\}  
\; \cdot  \; \pl{ \substack{a=1 \\ \not= p} }{r+2} \Bigg\{ \f{ \pl{b=r}{N-1} ( z_b-w_a-i\hbar)  }{ w_a - w_p }  \Bigg\} \\
\; =\; (i)^{r(r-1)}  \Big( \pl{a=r}{r+2}(-w_a) \Big)^{r+1} \cdot 
\pl{ a=1 }{r+2}     \Bigg\{ \f{\pl{b=r}{N-1} ( z_b-w_a-i\hbar)}{ \pl{b=1}{N}(y_b-w_a-i\hbar) } \Bigg\}  
  \; \cdot \; \mc{S}  \;. 
\end{multline}
There, we have set 
\beq
\mc{S} \; = \; \sul{p=1}{r+2} \; \pl{ \substack{a=1 \\ \not= p} }{r+2} \bigg\{ \f{1}{w_a - w_p}\bigg\} 
\; \cdot  \;  \pl{b=1}{N} ( y_b-w_p-i\hbar) \pl{b=r}{N-1} \bigg\{ \f{ 1 }{ z_b-w_p -i\hbar }  \bigg\} \;. 
\enq
This sum can be recast as a contour integral over the counter-clockwise loop $\msc{C}(\{w_a\}_1^{r+2})$ surrounding the points 
$\{w_a\}_1^{r+2}$ but not any other singularity of the integrand.
\beq
\mc{S} \; = \; - \Oint{ \msc{C}(\{w_a\}_1^{r+2}) }{}  \pl{a=1 }{r+2} \bigg\{ \f{1}{w_a - \tau }\bigg\} 
\; \cdot  \;  \pl{b=1}{N} ( y_b-\tau-i\hbar) \pl{b=r}{N-1} \bigg\{ \f{ 1 }{ z_b-\tau  -i\hbar }  \bigg\}  \cdot 
\f{\dd \tau }{2i\pi} \;. 
\enq

The integrand decays as $\tau^{-2}$ at infinity, so that there is no residue at $\infty$ and the only poles of the integrand
lying outside of $\msc{C}(\{w_a\}_1^{r+2})$ are at 
\beq
\tau \; = \; z_b -i\hbar  \quad \e{for} \quad b=r,\dots, N-1 \;. 
\enq
Thence, taking the integral by the residues lying outside of the contour of integration yields
\beq
\mc{S} \; = \; - \sul{p=r}{N-1} \pl{ \substack{ b=r \\ \not= p} }{ N-1 } \bigg\{ \f{ 1 }{ z_b - z_p  } \bigg\}
\; \cdot \; \f{ \pl{b=1}{N}(y_b - z_p) }{  \pl{a=1}{r+2} (w_a - z_p +i\hbar)   } \;. 
\enq
Taking the $w_a\tend \infty$, $a=r,\dots,r+2$, limit on the level of the last formula is straightforward. 
One ultimately gets that 
\beq
\lim_{\la \tend \infty}  \Big\{ \la^{r+1-N } t_{\la} (\bs{w} ,  \bs{z}  \mid \bs{y}_N)  \Big\}  \; = \; 
(i)^{r(r-1)} \pl{ a=1 }{r-1}   \bigg\{ \f{\pl{b=r}{N-1} ( z_b-w_a-i\hbar)}{ \pl{b=1}{N}(y_b-w_a-i\hbar) } \bigg\}  
\times 
\sul{p=r}{N-1} \pl{ \substack{ b=r \\ \not= p} }{ N-1 } \bigg\{ \f{ 1 }{ z_b - z_p  } \bigg\}
\f{ \pl{b=1}{N}(y_b - z_p) }{  \pl{a=1}{r-1} (w_a - z_p +i\hbar)   } \;. 
\enq

As a consequence, the leading asymptotics at $\la \tend \infty$ of the polynomials $t_{\la} (\bs{w} ,  \bs{z}  \mid \bs{y}) $ 
and $h_{\la} (\bs{w} ,  \bs{z}  \mid \bs{y}) $ coincide. Thus, we have provided the equality at enough interpolation points so
as to  ensure that the two polynomials are equal.
\qed


\begin{thebibliography}{10}
\bibitem{AnCompletenessEigenfunctionsTodaPeriodic}
D.~An, \emph{{"Complete Set of Eigenfunctions of the Quantum Toda Chain."}},
  Lett. Math. Phys. \textbf{\bf{87}} (2009), 209--223.

\bibitem{BabelonActionPositionOpsWhittakerFctions}
O.~Babelon, \emph{{"Equations in Dual Variables for Whittaker Functions."}},
  Lett.Math.Phys. \textbf{\bf{65}} (2003), 229--240.

\bibitem{BabelonQuantumInverseProblemConjClosedToda}
\bysame, \emph{{"On the Quantum Inverse Problem for the Closed Toda Chain."}},
  J.Phys.A \textbf{\bf{37}} (2004), 303--316.

\bibitem{BaxterPartitionfunction8Vertex-FreeEnergy}
R.J. Baxter, \emph{{"Partition function of the eight vertex lattice model."}},
  Ann. Phys. \textbf{\bf 70} (1972), 193--228.

\bibitem{DerkachovKorchemskyManashovXXXSoVandQopNewConstEigenfctsBOp}
S.~E. Derkachov, G.~P. Korchemsky, and A.~N. Manashov, \emph{{"Noncompact
  Heisenberg spin magnets from high-energy QCD: I. Baxter Q-operator and
  Separation of Variables."}}, Nucl. Phys. \textbf{\bf{B617}} (2001), 375--440.

\bibitem{FeherActionAngleMapOpenClassicalToda}
L.~Feh\'{e}r, \emph{{"Action-angle map and duality for the open Toda lattice in
  the perspective of Hamiltonian reduction."}}, Phys.Lett. \textbf{A377} (2013), 2917-2921.

\bibitem{FlaschkaLaxMatrixIntegrabilityClassicalToda}
H.~Flaschka, \emph{{"The Toda lattice II: Existence of integrals."}}, Phys.
  Rev. \textbf{\bf B 9} (1974), 1924--1925.

\bibitem{GaudinPasquierQOpConstructionForTodaChain}
M.~Gaudin and V.~Pasquier, \emph{{"The periodic Toda chain and a matrix
  generalization of the Bessel function recursion relations."}}, J. Phys. A:
  Math. Gen. \textbf{\bf 25} (1992), 5243--5252.

\bibitem{GerasimovKharchevLebedevRepThandQISM}
A.~Gerasimov, S.~Kharchev, and D.~Lebedev, \emph{{"Representation Theory and
  Quantum Inverse Scattering Method: The Open Toda Chain and the Hyperbolic
  Sutherland Model."}}, Int. Math. Res. Notices \textbf{\bf 17} (2004),
  823--854.

\bibitem{GiventalGaussGivIntRepObtainedForEFOfOpenToda}
A.~Givental, \emph{{"Stationary Phase Integrals, Quantum Toda Lattices, Flag
  Manifolds and the Mirror Conjecture."}}, AMS Trans. (2) \textbf{\bf 180}
  (1997), 103--115.

\bibitem{GoodmannWallachQuantumTodaIII}
R.~Goodman and N.R. Wallach, \emph{{"Classical and quantum-mechanical systems
  of Toda lattice type. III"}}, Comm. Math. Phys. \textbf{105} (1986),
  473--509.

\bibitem{GrosjeanMailletNiccoliFFofLatticeSineG}
N.~Grosjean, J.~M. Maillet, and G.~Niccoli, \emph{{"On the form factors of
  local operators in the lattice sine-Gordon model."}}, J. Stat. Mech.: Th. and
  Exp. (2012), P10006.

\bibitem{GutzwillerResolutionTodaChainSmallNPaper1}
M.~C. Gutzwiller, \emph{{"The quantum mechanical Toda lattice."}}, Ann. Phys.
  \textbf{\bf 124} (1980), 347--387.

\bibitem{GutzwillerResolutionTodaChainSmallNPaper2}
\bysame, \emph{{"The quantum mechanical Toda lattice II."}}, Ann. Phys.
  \textbf{\bf 133} (1981), 304--331.

\bibitem{IorgovShaduraIntRepEigenFctonsBopTodaWithBdry}
N.~Z. Iorgov and V.~N. Shadura, \emph{{"Wave functions of the Toda chain with
  boundary interaction."}}, Theor. Math. Phys. \textbf{\bf{142}} (2005),
  289--305.

\bibitem{KacMoerbeckeFullSolutionTodaAbelianIntegrals}
M.~Kac and P.~Van Moerbecke, \emph{{"A complete solution of the periodic Toda
  chain."}}, Proc. Nat. Acad. Sci. \textbf{\bf 72} (1975), 2879--2880.

\bibitem{KharchevLebedevMellinBarnesIntRepForWhittakerGLN}
S.~Kharchev and D.~Lebedev, \emph{{"Eigenfunctions of $GL(N,\mathbb{R})$ Toda
  chain: The Mellin-Barnes representation."}}, JETP Lett. \textbf{\bf 71}
  (2000), 235--238.

\bibitem{KharchevLebedevIntRepEigenfctsPeriodicTodaFromRecConstrofEigenFctOfB}
\bysame, \emph{{"Integral representations for the eigenfunctions of quantum
  open and periodic Toda chains from QISM formalism."}}, J.Phys.A \textbf{\bf
  34} (2001), 2247--2258.

\bibitem{KMTFormfactorsperiodicXXZ}
N.~Kitanine, J.-M. Maillet, and V.~Terras, \emph{{"Form factors of the XXZ
  Heisenberg spin-1/2 finite chain."}}, J. Phys. A: Math. Gen. \textbf{35}
  (2002), L753--10502.

\bibitem{KostantIdentificationOfEigenfunctionsOpenTodaAndWhittakerVectors}
B.~Kostant, \emph{{"Quantization and representation theory."}}, In
  "Representation theory of Lie groups", Proc. SCR/LMS res. symp. on rep. of
  Lie groups, London Math. Soc. Lect. Note \textbf{34} (1979), 287--316.

\bibitem{KostantIdentificationOfEigenfunctionsOpenTodaAndWhittakerVectorsMoreD%
eep}
\bysame, \emph{{"The solution to a generalized Toda lattice and representation
  theory."}}, Adv. in Math. \textbf{34} (1979), 195--338.

\bibitem{KozUnitarityofSoVTransform}
K.~K. Kozlowski, \emph{{"Unitarity of the SoV transform for the Toda chain."}},
  math.ph:1306.4967.

\bibitem{KozTeschnerTBAToda}
K.~K. Kozlowski and J.~Teschner, \emph{{"TBA for the Toda chain."}},
  Festschrift volume for Tetsuji Miwa, "Infinite Analysis 09: New Trends in
  Quantum Integrable Systems". (math-ph/10062906).

\bibitem{KuznetsovIPforclassicalSL(2)ChainsRecReconstrSepVars}
V.~B. Kuznetsov, \emph{{"Inverse problem for sl(2) lattices."}}, Proc. Int.
  Conf., Symmetry and Perturbation Theory, World Scientific (2002), 136--152.

\bibitem{MailletTerrasGeneralsolutionInverseProblem}
J.-M. Maillet and V.~Terras, \emph{{"On the quantum inverse scattering
  problem."}}, Nucl. Phys. B \textbf{\bf 575} (2000), 627--644.

\bibitem{NekrasovShatashviliConjectureTBADescriptionSpectrumIntModels}
N.~A. Nekrasov and S.~L. Shatashvili, \emph{{"Quantization of Integrable
  Systems and Four Dimensional Gauge Theories."}}, Proc. 16th Int. Congr. Math.
  Phys., Prague, Editor : P. Exner, World Scientific 2010 (2009), 265--289.

\bibitem{NiccoliCompleteSpectrumAndSomeFormFactorsInhomogeneousOpenXXZChain}
G.~Niccoli, \emph{{"Non-diagonal open spin-1/2 XXZ quantum chains by separation
  of variables: Complete spectrum and matrix elements of some quasi-local
  operators ."}}, J.Stat.Mech. (2012), P10025.

\bibitem{OlshanetskyPerelomovIntegralsOfMotionSemi-SimpleLieAlgebraSystems}
M.A. Olshanetsky and A.M. Perelomov, \emph{{"Quantum completely integrable
  systems connected with semi-simple Lie algebras."}}, Lett. Math. Phys.
  \textbf{\bf 2} (1977), 7--13.

\bibitem{OotaInverseProblemForFieldTheoriesIntegrability}
T.~Oota, \emph{{"Quantum projectors and local operators in lattice integrable
  models."}}, J. Phys. A: Math. Gen. \textbf{\bf 37} (2004), 441--452.

\bibitem{RuijsenaarsActionAngleMapsForVeriousSystems}
S.~Ruijsenaars, \emph{{"Action-angle maps and scattering theory for some
  finite-dimensional integrable systems III. Sutherland type systems and their
  duals."}}, Publ. Resch. Inst. Math. Sci. \textbf{\bf 31} (1995), 247--353.

\bibitem{SemenovTianShanskyQuantOpenTodaLatticesProofOrthogonalityFormulaForWh%
ittVectrs}
M.~A. Semenov-Tian-Shansky, \emph{{"Quantization of Open Toda Lattices."}},
  Encycl. Math. Sci., Vol {\bf 16 } Dynamical Systems VII, Edts V. I.~ Arnol'd
  and S. P.~Novikov, Springer-Verlag, Berlin (1994), 226--259.

\bibitem{SilantyevScalarProductFormulaTodaChain}
A.~V. Silantyev, \emph{{"Transition function for the Toda chain."}}, Theor.
  Math. Phys. \textbf{\bf 150} (2007), 315--331.

\bibitem{SklyaninResolutionIPFromQDet}
E.~K. Sklyanin, \emph{{"Bispectrality for the quantum open Toda chain."}},
  nlin.SI://1306.0454.

\bibitem{SklyaninSoVFirstIntroTodaChain}
\bysame, \emph{{"The quantum Toda chain."}}, Lect. Notes in Phys. \textbf{\bf
  226} (1985), 196--233.

\bibitem{SklyaninSoVGeneralOverviewAndConstrRecVectPofB}
\bysame, \emph{{"Quantum Inverse Scattering Method. Selected Topics."}}, in
  "Quantum Group and Quantum Integrable Systems" (Nankai Lectures in
  Mathematical Physics), ed. by Mo-Lin Ge, Singapore: World Scientific (1992),
  63--97.

\bibitem{TodaIntroTodaAndClassicalSolutionTodaChain}
W.~Toda, \emph{{"Wave propagation in anharmonic lattices."}}, J. Phys. Soc.
  Jap. \textbf{\bf{23}} (1967), 501--506.

\bibitem{WallachRealReductiveGroupsII}
N.~R. Wallach, \emph{{"Real reductive groups II."}}, Pure and applied
  mathematics, vol. 132-II, Academic Press, inc., 1992.
  
  \end{thebibliography}
\end{document}